\documentclass[prb,twocolumn,amssymb,amsmath,eqsecnum,%
floatfix]{revtex4}
\usepackage{bm,graphicx,dcolumn}
\begin{document}
\title{Consequences of local gauge symmetry in empirical tight-binding
theory}
\author{Bradley A. Foreman}
\email{phbaf@ust.hk}
\affiliation{Department of Physics and 
             Institute of Nano Science and Technology,
             The Hong Kong University of Science and Technology,
             Clear Water Bay, Kowloon, Hong~Kong, China}

\begin{abstract}
A method for incorporating electromagnetic fields into empirical
tight-binding theory is derived from the principle of local gauge
symmetry.  Gauge invariance is shown to be incompatible with empirical
tight-binding theory unless a representation exists in which the
coordinate operator is diagonal.  The present approach takes this
basis as fundamental and uses group theory to construct symmetrized
linear combinations of discrete coordinate eigenkets.  This produces
orthogonal atomic-like ``orbitals'' that may be used as a
tight-binding basis.  The coordinate matrix in the latter basis
includes intra-atomic matrix elements between different orbitals on
the same atom.  Lattice gauge theory is then used to define discrete
electromagnetic fields and their interaction with electrons.  Local
gauge symmetry is shown to impose strong restrictions limiting the
range of the Hamiltonian in the coordinate basis.  The theory is
applied to the semiconductors Ge and Si, for which it is shown that a
basis of 15 orbitals per atom provides a satisfactory description of
the valence bands and the lowest conduction bands.  Calculations of
the dielectric function demonstrate that this model yields an accurate
joint density of states, but underestimates the oscillator strength by
about 20\% in comparison to a nonlocal empirical pseudopotential
calculation.
\end{abstract}


\maketitle

\section{Introduction}

Tight-binding theory was originally proposed as an {\em ab initio}
technique for calculating the electronic properties of crystalline
solids from atomic wave functions.\cite{Bloch28} However,
first-principles calculations based on a linear combination of atomic
orbitals (LCAO) are computationally very demanding, and the
tight-binding approach met with relatively little success until Slater
and Koster suggested that it be used as an interpolation
scheme,\cite{SlaKos54} in which the Hamiltonian matrix elements are
fitted to experimental data or to band structures computed by other
methods.  This made it possible to describe atomic-level physics in a
basis of minimal size, leading to wide-ranging applications in many
areas of condensed-matter
physics. \cite{Har89,Har99,WangHo96,Bull80,Gor97,Turchi98} With modern
computer capabilities, first-principles electronic-structure
calculations are now commonplace, and {\em ab initio} tight-binding
theories are flourishing. \cite{Bull80,Gor97,Turchi98} Yet even today,
the empirical theory \cite{SlaKos54} predominates (even for the
fitting of first-principles calculations) because it is simple and
physically intuitive.

The formalism of \citet{SlaKos54} is incomplete, however, in that it
contains no prescription for coupling the electronic system to
external electromagnetic fields.  In {\em ab initio} theories,
\cite{Bull80,Gor97,Turchi98} one can use minimal coupling (with
suitable modifications for nonlocal potentials \cite{IsChLouie01}) and
calculate directly the necessary matrix elements of the momentum or
velocity operator.  In the empirical theory, these matrix elements can
simply be treated as extra fitting parameters,
\cite{ScCh85b,ChaAsp90,Xu90} determined by fitting the dielectric
function (and thus oscillator strengths) to experimental or
first-principles spectra.  However, even with the full use of symmetry
restrictions, the number of additional parameters can be undesirably
large; for example, \citet{ChaAsp90} have proposed an $sp^3d^2$ model
for GaAs with 13 Hamiltonian parameters and 17 independent momentum
parameters.

It is therefore clearly desirable to find ways of reducing or
eliminating these extra fitting parameters.  One possibility is to
define a kinematic momentum operator (equal to mass $m$ times
velocity) by
\begin{equation}
   \mathbf{p} = \frac{m}{i \hbar} [ \mathbf{x}, H ] , \label{eq:pxH}
\end{equation}
where \textbf{x} is the coordinate of the electron and $H$ is the
Hamiltonian.  In a sense this merely trades one problem for another,
since the coordinate matrix elements are still unknown, and the number
of these allowed by symmetry is no less than the number of momentum
matrix elements.  However, it is physically reasonable to simplify the
coordinate matrix by setting all nonlocal matrix elements to zero:
\begin{equation}
   \langle \alpha, \mathbf{x}_i | \mathbf{x} | \alpha',
   \mathbf{x}_{i'} \rangle = \delta_{i i'} [\delta_{\alpha \alpha'}
   \mathbf{x}_i + \mathbf{x}_{\alpha \alpha'} (i)] .  \label{eq:x}
\end{equation}
Here $| \alpha, \mathbf{x}_i \rangle$ is the ket vector for an
orthogonalized atomic orbital (L\"owdin orbital \cite{Lowd50,Alt91})
of type $\alpha$ located at position $\mathbf{x}_i$.  The parameter
$\mathbf{x}_{\alpha \alpha'}$ is an intra-atomic matrix element
coupling orbitals $\alpha$ and $\alpha'$ on the same atom.

The simplest choice of all is to set
\begin{equation}
   \mathbf{x}_{\alpha \alpha'} \equiv 0 ;  \label{eq:x0}
\end{equation}
in this model, there are no fitting parameters beyond those found in
the Hamiltonian.
\cite{Dres67,Smith79,BreyTej83,Har89,Har99,Koil91,Tser93,LYVRM93,Boyk99}
A closely related approach is the Peierls substitution,
\cite{Pei33,GrafVogl95,Dum98,Boyk01,IsChLouie01,BoykVogl02} in which
the zero-field Hamiltonian matrix $\langle \alpha, \mathbf{x}_i | H |
\alpha', \mathbf{x}_{i'} \rangle$ for a particle of charge $e$ is
replaced by
\begin{equation}
   \langle \alpha, \mathbf{x}_i | H | \alpha', \mathbf{x}_{i'} \rangle
   \exp \left( \frac{ie}{\hbar c}
   \int_{\mathbf{x}_{i'}}^{\mathbf{x}_{i}} \mathbf{A} \cdot \text{d}
   \mathbf{x} \right) + e \phi(\mathbf{x}_i) \delta_{i i'}
   \delta_{\alpha \alpha'} \label{eq:Peierls}
\end{equation}
in the presence of a vector potential $\mathbf{A}$ and scalar
potential $\phi$.  If the path of integration in Eq.\
(\ref{eq:Peierls}) is chosen to be a straight line,
\cite{GrafVogl95,Boyk01,IsChLouie01} then the linear term in the
Taylor series expansion of this equation is the same as the
$\mathbf{A} \cdot \mathbf{p}$ coupling obtained from Eqs.\
(\ref{eq:pxH})--(\ref{eq:x0}). \cite{GrafVogl95}

The total elimination of extra fitting parameters makes this model an
attractive one.  However, by eliminating the intra-atomic matrix
elements $\mathbf{x}_{\alpha \alpha'}$, one obtains a tight-binding
model that is not valid in the tight-binding limit of isolated atoms.
Thus, although the model should provide a reasonable description of
inter-atomic transitions between extended states, one has less
confidence in its ability to describe localized states, which may be
important at surfaces or interfaces.  Many authors have therefore
suggested augmenting the zero-parameter model by including a small
number of intra-atomic matrix elements.
\cite{Sel86,Rein94,BenVan96,LeuWha97,LeuWha98,Cruz99,PedKri01} It has
been shown for porous Si that, although the intra-atomic matrix
elements are small in magnitude (in a Bloch-function basis
\cite{Bloch28}), the {\em interference} between these terms and the
inter-atomic matrix elements contributes 25\% of the total
absorption. \cite{Cruz99} Thus, it appears that a quantitative
treatment of nanostructures may not be possible (in general) without
the inclusion of intra-atomic matrix elements.

The main difficulty with such models
\cite{Sel86,Rein94,BenVan96,LeuWha97,LeuWha98,Cruz99,PedKri01} is that
they are not gauge invariant. \cite{BoykVogl02} As shown by the
examples in Refs.\ \onlinecite{IsChLouie01} and \onlinecite{GovUng98},
lack of gauge invariance can lead to gross qualitative errors in the
predicted values of physical quantities.  Thus, there are significant
problems with both approaches considered above.  The models with
$\mathbf{x}_{\alpha \alpha'} = 0$ are gauge invariant, but they cannot
describe intra-atomic transitions.  The models with
$\mathbf{x}_{\alpha \alpha'} \neq 0$ can describe intra-atomic
transitions, but they are not gauge invariant.

The purpose of this paper is to demonstrate a technique for
constructing tight-binding models that are gauge invariant and provide
a full description of intra-atomic transitions.  This is achieved by
treating empirical tight-binding theory not as an approximation
derived from the Schr\"odinger equation, but as a fundamental
quantum-mechanical system in its own right.  This theory is required
to satisfy all of the basic principles of quantum mechanics, the most
important of which (in the present context) is the principle of local
gauge symmetry.
\cite{Weyl29,Dirac31,YangMills54,WuYang75,Mor83,Guidry91} The essence
of this principle is the concept that electromagnetism in quantum
mechanics is the gauge-invariant manifestation of a nonintegrable
(i.e., path-dependent) phase factor. \cite{Dirac31,WuYang75}

As will be shown in Sec.\ \ref{sec:coordinate} below, the reason why
existing models with intra-atomic coupling
\cite{Sel86,Rein94,BenVan96,LeuWha97,LeuWha98,Cruz99,PedKri01} are not
gauge invariant is that the coordinates $x$, $y$, and $z$ do not
commute. \cite{BoykVogl02} Gauge invariance requires commuting
coordinates, the existence of which implies the existence of a basis
of coordinate eigenkets.  Since empirical tight-binding theory deals
with finite vector spaces, the coordinate basis is necessarily
discrete.  Hence, the most general gauge-invariant finite vector space
is a set of discrete coordinate eigenkets.  This basis may be
transformed to a tight-binding basis by constructing ``orbitals'' from
symmetrized combinations of coordinate eigenkets (using well-known
techniques for symmetrizing plane waves \cite{Tink64,Bass66}).

The concept of gauge symmetry on a discrete lattice is not new, having
appeared many years ago as a technique for imposing a momentum cutoff
in quantum chromodynamics.
\cite{Guidry91,Wil74,KogSus75,Rebbi83,Rothe97} \citet{GovUng98} have
recently suggested using lattice gauge techniques in empirical
tight-binding theory.  However, their proposal, like most applications
of lattice gauge theory, is based on a simple cubic lattice.  As shown
below, the simple cubic lattice is unsuitable for practical
tight-binding models because it can only achieve sufficient accuracy
with an unreasonably large basis (i.e., a very small lattice
constant).  Thus, the development of efficient tight-binding models
requires consideration of more general geometries.

\citet*{ChrFriLee82a,ChrFriLee82b,ChrFriLee82c} have developed a
lattice gauge theory for {\em random} lattices, which (with some
slight modifications) is sufficiently general for the present
purposes.  However, the complete formal machinery of quantum
chromodynamics is somewhat cumbersome when one is dealing only with
simple electromagnetism.  Thus, for reasons of clarity, the author has
chosen to present the theory in terms of a simple but elegant approach
used by Dirac. \cite{Dirac31} After a preliminary discussion of
topology (i.e., how an electron is permitted to move from one lattice
site to another) in Sec.\ \ref{sec:topology}, Sec.\ \ref{sec:gauge}
presents an adaptation of Dirac's analysis \cite{Dirac31} to the case
of a discrete lattice.  The outcome is a gauge-invariant formulation
of electromagnetism in empirical tight-binding theory.

Although the theory derived in this way has many similarities with
conventional tight-binding theory, there are significant differences
as well.  Not all tight-binding models can be made gauge invariant;
\cite{note:tbgauge} this is possible only if the basis can be
constructed from symmetrized coordinate eigenkets.  In addition, local
gauge symmetry imposes strong restrictions on the Hamiltonian matrix,
which have the effect of sharply reducing the number of allowed
Hamiltonian fitting parameters.  Finally, unlike previous empirical
tight-binding theories, the present approach provides an explicit
(discrete) wave function for the electron.

The formalism derived here is applied to two semiconductors with the
diamond structure (Ge and Si) in Sec.\ \ref{sec:application}.  For
these systems, a basis of 15 orbitals per atom is shown to provide a
satisfactory fit to the valence bands and the lowest conduction bands
(up to about 5~eV above the valence-band maximum).  These results are
comparable to those obtained from a 10-orbital basis proposed recently
in Ref.\ \onlinecite{Jancu98}.  The basis used here is 50\% larger,
but their model \cite{Jancu98} cannot be made gauge invariant if
intra-atomic coupling is included.  Thus, it appears that some
tradeoffs are necessary if gauge invariance is to be achieved.

\section{Coordinate matrices and the coordinate representation}

\label{sec:coordinate}

As mentioned above, the intra-atomic coupling used in existing
tight-binding models
\cite{Sel86,Rein94,BenVan96,LeuWha97,LeuWha98,Cruz99,PedKri01} leads
to a lack of gauge invariance. \cite{BoykVogl02} This may be seen from
a simple $sp^3$ model for a single atom.  In this case, we know from
atomic physics that there are coordinate matrix elements coupling the
$s$ and $p$ orbitals:
\begin{equation}
   \langle s | x | p_x \rangle = \langle s | y | p_y \rangle = \langle
   s | z | p_z \rangle \equiv c ,
\end{equation}
where $c$ is real.  In the basis $\{ |s\rangle, |p_x\rangle,
|p_y\rangle, |p_z\rangle \}$, the matrices representing $x$ and $y$
are therefore
\begin{equation}
   x = \left[
   \begin{array}{cccc}
      0 & c & 0 & 0 \\
      c & 0 & 0 & 0 \\
      0 & 0 & 0 & 0 \\
      0 & 0 & 0 & 0
   \end{array}
   \right] , \qquad y = \left[
   \begin{array}{cccc}
      0 & 0 & c & 0 \\
      0 & 0 & 0 & 0 \\
      c & 0 & 0 & 0 \\
      0 & 0 & 0 & 0
   \end{array}
   \right] .  \label{eq:xy1}
\end{equation}
But this implies that $x$ and $y$ do not commute:
\begin{equation}
   xy - yx = \left[
   \begin{array}{cccc}
      0 & 0 & 0 & 0 \\
      0 & 0 & c^2 & 0 \\
      0 & -c^2 & 0 & 0 \\
      0 & 0 & 0 & 0
   \end{array}
   \right] \neq 0 .
\end{equation}
This means that the coordinate representation (consisting of
simultaneous eigenkets of $x$, $y$, and $z$) does not exist.  Even
more important, it means that the theory cannot be gauge invariant.
In a gauge transformation, the vector and scalar potentials transform
as
\begin{equation}
   \mathbf{A} \rightarrow \mathbf{A} + \nabla \Lambda , \qquad \phi
   \rightarrow \phi - \frac{1}{c} \frac{\partial \Lambda}{\partial t} ,
\end{equation}
and the state ket $| \psi \rangle$ transforms as
\begin{equation}
   | \psi \rangle \rightarrow U | \psi \rangle, \qquad U \equiv \exp
   \left[ \frac{i e \Lambda (\mathbf{x},t)}{\hbar c} \right] ,
\end{equation}
where $\Lambda$ is an arbitrary function of $\mathbf{x} = (x,y,z)$ and
$t$.  If a theory is gauge invariant, all physically measurable
quantities must be independent of such transformations.  But the
expectation value $\langle x \rangle$ is a measurable quantity, and
under a gauge transformation one has
\begin{equation}
   \langle x \rangle \rightarrow \langle U^{\dagger} x U \rangle ,
\end{equation}
where $U^{\dagger} x U \neq x$ if $\Lambda$ depends on $y$ or $z$.
Hence, no theory can be gauge invariant if $x$, $y$, and $z$ do not
commute.

Is there any way of achieving gauge invariance without setting $c =
0$?  Perhaps the $sp^3$ basis is too small, and the situation might be
improved by including more orbitals ($d,f,\ldots$).  However, one soon
finds that for any {\em finite} LCAO basis, the lack of gauge
invariance persists.  This follows directly from the Wigner-Eckart
theorem---since $\mathbf{x}$ is a vector operator, it couples states
with angular momentum $l$ to those with $l \pm 1$.  Hence, any finite
truncation of the basis results in non-commuting coordinates.

Another possibility is to keep the same $sp^3$ basis, but modify the
coordinate matrix.  The physical justification for doing so is the
fact that the orbitals used in empirical tight-binding theory are not
atomic orbitals; they are {\em orthogonalized} atomic
orbitals. \cite{Lowd50} Therefore, they do not have the full
rotational symmetry of atomic orbitals; they have only the site
symmetry of the crystal structure.  For example, the atoms in a
diamond crystal have site symmetry $T_d$.  \cite{Hahn92} Therefore,
the orbital that was denoted $|p_z\rangle$ above should really be
written as $|\Gamma_{15}^{z} \rangle$, since it belongs to the
$\Gamma_{15}$ representation of $T_d$. \cite{Parm55b}

However, the $d$ orbital $| d_{xy} \rangle$ also transforms as
$|\Gamma_{15}^{z} \rangle$.  Thus, in the $T_d$ group, the matrix element
$b \equiv \langle \Gamma_{15}^{x} | y | \Gamma_{15}^{z} \rangle$ is allowed,
and the coordinate matrices (\ref{eq:xy1}) become
\begin{equation}
   x = \left[
   \begin{array}{cccc}
      0 & c & 0 & 0 \\
      c & 0 & 0 & 0 \\
      0 & 0 & 0 & b \\
      0 & 0 & b & 0
   \end{array}
   \right] , \qquad y = \left[
   \begin{array}{cccc}
      0 & 0 & c & 0 \\
      0 & 0 & 0 & b \\
      c & 0 & 0 & 0 \\
      0 & b & 0 & 0
   \end{array}
   \right] .
\end{equation}
This yields
\begin{equation}
   xy - yx = \left[
   \begin{array}{cccc}
      0 & 0 & 0 & 0 \\
      0 & 0 & c^2 - b^2 & 0 \\
      0 & b^2 - c^2 & 0 & 0 \\
      0 & 0 & 0 & 0
   \end{array}
   \right] ,
\end{equation}
which is equal to zero if $b = \pm c$.

Setting $b = c$ would imply that the $p$ and $d$ orbitals have equal
weight in the $\Gamma_{15}$ states.  This is not as absurd as it
sounds; \citet{BogGor94} have shown using first-principles
pseudopotential calculations that for the $\Gamma_{15}$ states at the
top of the valence band in GaAs, the probability of finding an
electron in a cation $d$ orbital is {\em greater} than that of finding
it in a cation $p$ orbital.  (In AlAs, the probability ratio is
greater than 2.\cite{BogGor94}) Thus, it is not unreasonable to assume
that $b$ and $c$ have comparable magnitudes.

If one sets $b = c$, then the coordinate operators have simultaneous
eigenkets $| x', y', z' \rangle$, which are given by
\begin{eqnarray}
   | ccc \rangle & = & \textstyle\frac12 ( | \Gamma_{1} \rangle + |
     \Gamma_{15}^{x} \rangle + | \Gamma_{15}^{y} \rangle + | \Gamma_{15}^{z}
     \rangle ) , \nonumber \\
   | c\bar{c}\bar{c} \rangle & = & \textstyle\frac12 ( | \Gamma_{1} \rangle
     + | \Gamma_{15}^{x} \rangle - | \Gamma_{15}^{y} \rangle - |
     \Gamma_{15}^{z} \rangle ) , \nonumber \\
   | \bar{c}c\bar{c} \rangle & = & \textstyle\frac12 ( | \Gamma_{1} \rangle
     - | \Gamma_{15}^{x} \rangle + | \Gamma_{15}^{y} \rangle - |
     \Gamma_{15}^{z} \rangle ) , \nonumber \\
   | \bar{c}\bar{c}c \rangle & = & \textstyle\frac12 ( | \Gamma_{1} \rangle
     - | \Gamma_{15}^{x} \rangle - | \Gamma_{15}^{y} \rangle + |
     \Gamma_{15}^{z} \rangle ) .  \label{eq:ccc}
\end{eqnarray}
Note that the coordinate eigenvalues are located at the corners of a
tetrahedron.  In fact, the linear combinations given in (\ref{eq:ccc})
are identical to the hybrid bond orbitals used in analytical
tight-binding theories, \cite{Har89,Har99} although these are not
ordinarily interpreted as exact coordinate eigenkets because the
$\Gamma_{15}$ states are assumed to be pure $p$ orbitals.

The procedure outlined above is rather clumsy; one simply modifies the
coordinate matrices by trial and error in an attempt to make them
commute.  One cannot predict in advance whether the attempt will
succeed, and in general it will not.  However, the unitary
transformation (\ref{eq:ccc}) may be inverted to obtain
\begin{eqnarray}
   | \Gamma_{1} \rangle & = & \textstyle\frac12 ( | ccc \rangle + | c
     \bar{c} \bar{c} \rangle + | \bar{c} c \bar{c} \rangle + | \bar{c}
     \bar{c} c \rangle ) , \nonumber \\
   | \Gamma_{15}^{x} \rangle & = & \textstyle\frac12 ( | ccc \rangle + | c
     \bar{c} \bar{c} \rangle - | \bar{c} c \bar{c} \rangle - | \bar{c}
     \bar{c} c \rangle ) , \nonumber \\
   | \Gamma_{15}^{y} \rangle & = & \textstyle\frac12 ( | ccc \rangle - | c
     \bar{c} \bar{c} \rangle + | \bar{c} c \bar{c} \rangle - | \bar{c}
     \bar{c} c \rangle ) , \nonumber \\
   | \Gamma_{15}^{z} \rangle & = & \textstyle\frac12 ( | ccc \rangle - | c
     \bar{c} \bar{c} \rangle - | \bar{c} c \bar{c} \rangle + | \bar{c}
     \bar{c} c \rangle ) , \label{eq:Td}
\end{eqnarray}
which immediately suggests a more fruitful approach.  The linear
combinations given in Eq.\ (\ref{eq:Td}) are just what one would
obtain by starting with a single coordinate eigenket (say
$|ccc\rangle$) and using the symmetry operations of the tetrahedral
group $T_d$ to construct ``symmetrized'' orbitals \cite{Tink64,Bass66}
that transform according to the irreducible representations of
$T_d$. \cite{Parm55b}

Thus, in this alternative approach, the coordinate basis is taken as
fundamental, and the tight-binding basis is merely a secondary
alternative that is useful for reasons of symmetry.  Since the
existence of a coordinate representation is necessary for gauge
invariance, no tight-binding basis can be made gauge invariant if it
cannot be represented in terms of symmetrized coordinate
eigenkets. \cite{note:tbgauge} Hence, this symmetrization procedure
provides us with all possible gauge-invariant tight-binding models.

The orbitals in Eq.\ (\ref{eq:Td}) are useful as a starting point, but
they cannot be interpreted as atomic orbitals because they do not have
inversion symmetry.  To obtain more ``atomic-like'' orbitals, one can
apply the symmetry operations of the cubic group $O_h$ to the basis
ket $| ccc \rangle$, which yields the orbitals
\begin{eqnarray}
   | \Gamma_1 \rangle & = & | s \rangle \nonumber \\ & = &
   \frac{1}{\sqrt{8}} ( | ccc \rangle + | c \bar{c} \bar{c} \rangle +
   | \bar{c} c \bar{c} \rangle + | \bar{c} \bar{c} c \rangle \nonumber
   \\ & & \qquad + | \bar{c}\bar{c}\bar{c} \rangle + | \bar{c}cc
   \rangle + | c\bar{c}c \rangle + | cc\bar{c} \rangle ) , \nonumber \\
   | \Gamma_{2'} \rangle & = & | f_{xyz} \rangle \nonumber \\ & =
   & \frac{1}{\sqrt{8}} ( | ccc \rangle + | c \bar{c} \bar{c} \rangle
   + | \bar{c} c \bar{c} \rangle + | \bar{c} \bar{c} c \rangle
   \nonumber \\ & & \qquad - | \bar{c}\bar{c}\bar{c} \rangle - |
   \bar{c}cc \rangle - | c\bar{c}c \rangle - | cc\bar{c} \rangle ) ,
   \nonumber \\
   | \Gamma_{15}^{z} \rangle & = & | p_z \rangle \nonumber \\ & = &
   \frac{1}{\sqrt{8}} ( | ccc \rangle - | c \bar{c} \bar{c} \rangle -
   | \bar{c} c \bar{c} \rangle + | \bar{c} \bar{c} c \rangle \nonumber
   \\ & & \qquad - | \bar{c}\bar{c}\bar{c} \rangle + | \bar{c}cc
   \rangle + | c\bar{c}c \rangle - | cc\bar{c} \rangle ) , \nonumber \\
   | {\Gamma}_{25'}^{xy} \rangle & = & | d_{xy} \rangle
   \label{eq:ccc_tb} \\
   & = & \frac{1}{\sqrt{8}} ( | ccc \rangle - | c \bar{c} \bar{c}
   \rangle - | \bar{c} c \bar{c} \rangle + | \bar{c} \bar{c} c \rangle
   \nonumber \\ & & \qquad + | \bar{c}\bar{c}\bar{c} \rangle - |
   \bar{c}cc \rangle - | c\bar{c}c \rangle + | cc\bar{c} \rangle ) .
    \nonumber
\end{eqnarray}
Here two different labels are used: the representations of $O_h$,
\cite{BSW36} and the conventional atomic orbital notation.  Other
orbitals not given here may be obtained from cyclic permutations of
$x$, $y$, and $z$.  Note that the $T_d$ orbital $|\Gamma_{15}^{z}
\rangle$ in Eq.\ (\ref{eq:Td}) is the same as $\frac{1}{\sqrt{2}}
(|p_z \rangle + |d_{xy}\rangle)$, whereas the $T_d$ orbital
$|\Gamma_1\rangle$ is just $\frac{1}{\sqrt{2}} (| s \rangle +
|f_{xyz}\rangle)$.

In the basis (\ref{eq:ccc_tb}), the nonzero coordinate matrix elements
are
\begin{equation}
   \langle s | x | p_x \rangle = \langle p_x | y | d_{xy} \rangle =
   \langle d_{xy} | z | f_{xyz} \rangle = c , \label{eq:x111}
\end{equation}
plus others given by cyclic permutations.  The selection rules for
$\mathbf{x}$ are thus the same as those in a spherically symmetric
atom---although, in any real atom, the matrix elements (\ref{eq:x111})
would not be numerically equal.  This equality occurs because the
basis kets (\ref{eq:ccc_tb}) are degenerate eigenkets of the radial
coordinate $r = \sqrt{x^2 + y^2 + z^2}$.  To break the numerical
equality, one would need to use basis functions with a linear
combination of different radii.

The above procedure may, of course, be applied to coordinate eigenkets
other than $|ccc\rangle$.  Below is a list of the representations
obtained by applying the symmetry operations of $O_h$ to several
different ``generator'' eigenkets:
\begin{eqnarray}
   | 000 \rangle & \rightarrow & \Gamma_1 
     \nonumber \\ & \rightarrow & s , \nonumber \\
   | 100 \rangle & \rightarrow & \Gamma_1 + \Gamma_{15} + \Gamma_{12}
     \nonumber \\ & \rightarrow & s^1p^3d^2 , \nonumber \\
   | 111 \rangle & \rightarrow & \Gamma_1 + \Gamma_{15} + \Gamma_{25'}
     + \Gamma_{2'} \nonumber \\ & \rightarrow & s^1p^3d^3f^1 , \nonumber \\
   | 110 \rangle & \rightarrow & \Gamma_1 + \Gamma_{12} + \Gamma_{15}
     + \Gamma_{25'} + \Gamma_{25} \nonumber \\ & \rightarrow &
     s^1p^3d^5f^3 . \label{eq:symm_basis}
\end{eqnarray}
Explicit basis functions for these representations are given in
Appendix \ref{app:basis}.

With these results, one can now construct a gauge-invariant
tight-binding model simply by putting a set (or more than one set) of
these ``orbitals'' on each atom in a crystal or molecule.  In such a
model, the coordinate operators $x$, $y$, and $z$ commute by
construction.  However, one is no longer permitted to choose orbitals
arbitrarily.  The choices are limited to taking {\em all} of the
orbitals in a set or taking {\em none} of them.  As an example, one
cannot discard the $f$ orbitals in the basis generated by
$|110\rangle$ without destroying the gauge invariance of the theory.

This approach yields a tight-binding model with orthogonal orbitals.
Another approach is to define a grid of coordinates, some points of
which are not uniquely associated with individual atoms.  One may
still construct symmetrized orbitals in this case, but the orbitals
are not orthogonal.  This makes the tight-binding approach more
difficult; however, one can simplify the theory by choosing a Bravais
lattice for the coordinate grid, in which case the model may be viewed
as a discrete pseudopotential model.  Applications of both the
pseudopotential and tight-binding approaches are considered below in
Sec.\ \ref{sec:application}.

\section{Topology of the lattice}

\label{sec:topology}

As we have seen, the most general gauge-invariant tight-binding basis
consists of a set of discrete coordinate eigenkets, which will be
referred to as a lattice.  Such a lattice is generally not periodic.
In order to apply the principle of local gauge symmetry to such a
system, one must be able to calculate the change in phase that occurs
along any specified path in coordinate space. \cite{Dirac31,WuYang75}
Thus, the first step is to define what is meant by a ``path'' in a
discrete coordinate system.

In general, a path is just an ordered sequence of points.  In a
continuous coordinate system, neighboring points in the sequence must
be separated by an infinitesimal distance.  This defines the {\em
topology} of the system, in which points are linked together only if
they are adjacent in coordinate space.  It is desirable to define the
topology of the discrete lattice in a similar way.

One way to do this is to construct a Voronoi polyhedron around each
site in the lattice. \cite{ChrFriLee82a} A Voronoi polyhedron is just
the region in space closest to that point; \cite{Okabe00} if the
lattice is a Bravais lattice, the polyhedron is the same as a
Wigner-Seitz cell.  In mathematical terms, the Voronoi polyhedron
$\Omega_i$ for site $\mathbf{x}_i$ is the set of points $\mathbf{x}$
such that $|\mathbf{x} - \mathbf{x}_i| \leq |\mathbf{x} -
\mathbf{x}_j|$ for all $j \neq i$.  The topology is then defined by
the following rule: If $\Omega_i$ and $\Omega_j$ share a surface with
area $S_{ij} > 0$, the sites $\mathbf{x}_i$ and $\mathbf{x}_j$ are
linked together; otherwise, they are not.

In some cases, it may happen that two Voronoi polyhedra share only a
point or a line, in which case $S_{ij} = 0$.  The linking algorithm
presented in Ref.\ \onlinecite{ChrFriLee82a} does not consider this
possibility (because Ref.\ \onlinecite{ChrFriLee82a} deals only with
random lattices, for which the probability of such an event is zero).
For certain applications, it is useful to include links between such
sites, \cite{Okabe00} but we shall see below that these links should
be excluded in the present situation.  Thus, only adjacent sites whose
Voronoi polyhedra share a surface with $S_{ij} > 0$ are linked.

A path in the discrete lattice is then just an ordered sequence of
linked points, and a closed path is one whose first and last points
are the same.  By definition, every edge of a Voronoi polyhedron is
equidistant from three or more lattice sites, all of which lie in a
plane perpendicular to the given edge.  These sites are closer to this
edge than any other sites.  The links between these sites form a
closed path, and the area bordered by the links is called a
``plaquette.''  There is a one-to-one relationship between the
plaquettes and the edges of the Voronoi polyhedra.

The plaquettes partition all of coordinate space into nonoverlapping
volumes called (Delaunay) cells.  Each cell is uniquely associated
with one corner of a Voronoi polyhedron.  The partitioning of space
into cells is referred to as a Delaunay tessellation. \cite{Okabe00}

A general algorithm for calculating the geometry of Voronoi polyhedra,
links, plaquettes, and cells is presented in Appendix
\ref{app:geometry}.  The expressions derived there will be of use in
what follows.

\section{Local gauge symmetry on an arbitrary discrete lattice}

\label{sec:gauge}

\citet*{ChrFriLee82a,ChrFriLee82b,ChrFriLee82c} have developed a
theory of local gauge symmetry on a random lattice.  This section
presents a modified version of their theory, with special emphasis on
the implications of the principle of local gauge symmetry for
tight-binding theory.  The presentation follows Dirac's approach,
\cite{Dirac31} in which the existence of electromagnetic fields is
``derived'' as a straightforward consequence of a degree of freedom
(nonintegrable phases) possessed by any quantum-mechanical system that
can be represented in a coordinate basis.

\subsection{Electromagnetism is a nonintegrable phase}

In a discrete coordinate basis, any ket vector may be expressed as
\begin{equation}
   | \psi \rangle = \sum_{i} c_{i} | \mathbf{x}_{i} \rangle ,
\end{equation}
where $| \mathbf{x}_{i} \rangle$ is a coordinate eigenket, which is
assumed to be normalized such that
\begin{equation}
   \langle \mathbf{x}_{i} | \mathbf{x}_{j} \rangle = \delta_{ij}
   . \label{eq:deltaij}
\end{equation}
Dirac's starting point \cite{Dirac31} is the fact that physical
predictions in quantum mechanics are ultimately expressed in terms of
probabilities of the form $| \langle \psi | \psi' \rangle |^2$, where
the probability amplitude $\langle \psi | \psi' \rangle$ is given by
\begin{equation}
   \langle \psi | \psi ' \rangle = \sum_i c_i^* c_i' .
\end{equation}
The probability is obviously well defined even when the overall phase
of $|\psi\rangle$ has no definite value.  This degree of freedom is
referred to as global gauge symmetry.

The existence of global gauge symmetry raises the question of whether
it is necessary for the local probability amplitude $c_i$ to have a
definite phase.  In other words, suppose we write
\begin{equation}
   c_i = b_i e^{i \beta_i} , \label{eq:phase}
\end{equation}
where the phase of $b_i$ is well defined (to within an integer
multiple of $2\pi$), but $\beta_i$ is a nonintegrable function---that
is, the change in $\beta_i$ around a closed path can take on any
value.  In this case one can see that $| \langle \psi | \psi' \rangle
|^2$ is well defined {\em only} if the change in $\beta_i$ around any
closed path is the {\em same} for all states $| \psi \rangle$ and $|
\psi' \rangle$ (to within an integer multiple of $2\pi$, which is
absorbed into the definition of $b_i$).  But anything that is the same
for all states can be viewed as a physically real part of the
dynamical system.  Since the present system consists only of a single
point particle, these nonintegrable phases must represent a field of
force acting on the particle.

The principle of local gauge symmetry is therefore defined by the
following two postulates: \cite{Dirac31} (i)~The physical predictions
of the theory must be unambiguous.  (ii)~The phase of $c_i$ at any
point in space and time need not be well defined; only the {\em
change} in phase between {\em linked} points must be definite.  As
shown above, these postulates entail that the change in $\beta_i$
around any closed path must be the same for all states.  According to
postulate (ii), this change is fixed for any path by the change in
$\beta_i$ between two linked points in space:
\begin{equation}
   \kappa_{ij} = \beta_i - \beta_{j} \qquad (i \text{ linked to } j) ,
   \label{eq:kappa}
\end{equation}
and in time:
\begin{equation}
   \lambda_i = \frac{\text{d} \beta_i}{\text{d} t} \equiv
   \dot{\beta}_i .
\end{equation}
Since $\beta_i$ is nonintegrable, $\kappa_{ij}$ and $\lambda_i$ are
{\em independent} variables.  These two quantities are the fundamental
dynamical variables that arise from the principle of local gauge
symmetry.  It will now be shown that $\kappa_{ij}$ and $\lambda_i$
can be interpreted as {\em potentials} for the electromagnetic field.

One possible closed path involves a space displacement
$\mathbf{d}_{ij} = \mathbf{x}_i - \mathbf{x}_j$ and an infinitesimal
time displacement $\text{d}t$, followed by $\mathbf{d}_{ji}$ and
$-\text{d}t$.  The change in phase around this path can be used to
define (tentatively) an electric field variable
\begin{subequations}
\label{eq:E}
\begin{equation}
   E_{ji} = - \frac{\hbar}{e} \frac{\lambda_{j} - \lambda_{i} -
   \dot{\kappa}_{ji}}{d_{ji}} , \label{eq:Eii}
\end{equation}
where $d_{ji} = | \mathbf{d}_{ji} |$.  If the index $\ell$ is used to
label the links, one may write this in the simpler form
\begin{equation}
   E_{\ell} = - \frac{\hbar}{e} \frac{\Delta \lambda_{\ell} -
   \dot{\kappa}_{\ell}}{d_{\ell}} . \label{eq:Ej}
\end{equation}
\end{subequations}
Here $E_{\ell}$ is interpreted as the component of the electric field
in the direction $\mathbf{d}_{\ell} = \mathbf{d}_{ji}$; the components
perpendicular to $\mathbf{d}_{\ell}$ are not defined.  Equation
(\ref{eq:E}) takes a familiar form if expressed in terms of the
potentials
\begin{equation}
   \phi_i = \frac{\hbar}{e} \lambda_i , \qquad A_{\ell} = -
   \frac{\hbar c}{e} \frac{\kappa_{\ell}}{d_{\ell}} ,
   \label{eq:potentials}
\end{equation}
since then
\begin{equation}
   E_{\ell} = - \frac{\Delta \phi_{\ell}}{d_{\ell}} - \frac{1}{c}
   \frac{\partial A_{\ell}}{\partial t} .
\end{equation}
Here the notation $\partial A_{\ell} / \partial t$ indicates that
$d_{\ell}$ is to be held constant during the differentiation.

Another type of closed path is an elementary plaquette $q$ constructed
from links $\ell$ in coordinate space (see Sec.\ \ref{sec:topology}
and Appendix \ref{app:geometry}).  The change in phase around the
perimeter of the plaquette may be used to define the magnetic field
\begin{equation}
   B_q = - \frac{1}{S_q} \frac{\hbar c}{e} \sum_{\ell \in q}
   \kappa_{\ell} = \frac{1}{S_q} \sum_{\ell \in q} A_{\ell} d_{\ell} ,
   \label{eq:B}
\end{equation}
where $S_q$ is the area of the plaquette [see Eq.\ (\ref{eq:Sq})].
Summing Eq.\ (\ref{eq:B}) over the (closed) surface of a cell $c$
leads immediately to the ``no monopoles'' law: \cite{note:monopoles}
\begin{equation}
   \sum_{q \in c} B_q S_q = 0 .
\end{equation}
Likewise, summing $E_{\ell} d_{\ell}$ around the perimeter of a
plaquette gives Faraday's law:
\begin{equation}
   \sum_{d_{\ell} \in S_q} E_{\ell} d_{\ell} = - \frac{1}{c}
   \frac{\text{d}}{\text{d}t}(B_q S_q) .
\end{equation}

The other two Maxwell equations can be obtained from the Lagrangian $L
= L_f + L_e$, where $L_f$ is the electromagnetic field Lagrangian
\begin{equation}
   L_f = \frac{1}{8 \pi} \sum_{\ell} 3 E_{\ell}^2 \Omega_{\ell} -
   \frac{1}{8 \pi} \sum_q 3 B_q^2 \Omega_q . \label{eq:Lf}
\end{equation}
Here $\Omega_{\ell} = \frac13 S_{\ell} d_{\ell}$ is the volume of link
$\ell$, where $S_{\ell} = S_{ij}$ is the area of the surface shared by
the Voronoi polyhedra for sites $i$ and $j$ (see Appendix
\ref{app:geometry}).  Likewise, $\Omega_q = \frac13 S_q d_q$ is the
volume of plaquette $q$, where $d_q$ is the length of the Voronoi
polyhedron edge corresponding to $q$.  Equation (\ref{eq:Lf}) is just
a discrete version of the standard field Lagrangian \cite{Jackson75}
$\frac{1}{8\pi}\int (E^2 - B^2) \text{d}^3 \! x$; the only apparent
difference is an extra factor of 3.  This factor cancels the factor of
$\frac13$ in the definition of $\Omega_{\ell}$ and $\Omega_q$, thus
leading to the correct Maxwell equations below.  It appears in Eq.\
(\ref{eq:Lf}) because the standard Lagrangian is expressed in terms of
$E^2 = E_x^2 + E_y^2 + E_z^2$, whereas $E_{\ell}^2$ includes only the
component of $\mathbf{E}$ in the direction of $\mathbf{d}_{\ell}$.

The electronic term in the Lagrangian, which includes the
field-particle coupling, is
\begin{subequations}
\begin{eqnarray}
   L_e & = & \frac{i \hbar}{2} \sum_i (c_i^* \dot{c}_i - \dot{c}_i^*
   c_i) - \sum_{i,j} c_i^* H_{ij} c_{j} \label{eq:Lea} \\ & = &
   \frac{i \hbar}{2} \sum_i (b_i^* \dot{b}_i - \dot{b}_i^* b_i) -
   \sum_{i,j} b_i^* \tilde{H}_{ij} b_{j} . \label{eq:Leb}
\end{eqnarray}
\end{subequations}
Here $H_{ij} = \langle \mathbf{x}_i | H | \mathbf{x}_{j} \rangle$ is
the Hamiltonian in the absence of electromagnetic fields, while
\begin{equation}
   \tilde{H}_{ij} = H_{ij} e^{-i (\beta_i - \beta_{j})} + \hbar
   \dot{\beta}_i \delta_{ij} . \label{eq:Htilde}
\end{equation}
The first expression (\ref{eq:Lea}) for $L_e$ has exactly the same
form as the Lagrangian in the case of no electromagnetic fields.  This
expresses the fundamental physical content of the principle of local
gauge symmetry---that the influence of the field upon the particle can
be expressed {\em entirely} in terms of the nonintegrable phase of the
probability amplitude $c_i = b_i e^{i \beta_i}$.  In the second
expression (\ref{eq:Leb}) for $L_e$, all of the nonintegrable phases
are collected together in the effective Hamiltonian $\tilde{H}_{ij}$.
This is the usual approach, in which the probability amplitude $b_i$
has a well-defined phase, and the field appears only in the
Hamiltonian.

The Hamiltonian (\ref{eq:Htilde}) appearing in the Lagrangian
(\ref{eq:Leb}) depends upon the phase difference $\beta_i - \beta_j$.
This phase difference is not well defined unless the sites $i$ and $j$
are linked.  But according to postulate (i) above, all physical
predictions of the theory must be unambiguous.  Hence, the principle
of local gauge symmetry demands that
\begin{equation}
   H_{ij} = 0 \qquad (i \text{ not linked to } j) . \label{eq:Hij0}
\end{equation}
Local gauge symmetry therefore imposes constraints not found in
conventional tight-binding models.

Note that the Lagrangian $L$ is gauge invariant by construction.  In
other words, both $L_f$ and $L_e$ are invariant under the gauge
(phase) transformation
\begin{eqnarray}
   b_i & \rightarrow & b_i e^{-i \chi_i} , \nonumber \\
   \lambda_i & \rightarrow & \lambda_i + \dot{\chi}_i , \\
   \kappa_{ij} & \rightarrow & \kappa_{ij} + \chi_i - \chi_{j} ,
   \nonumber
\end{eqnarray}
where $\chi_i$ is an arbitrary integrable function.

Given the above Lagrangian, the Euler-Lagrange equation for
$\lambda_i$ or $\phi_i$ is just Gauss's law
\begin{equation}
   \sum_{j} E_{ji} S_{ji} = 4 \pi q_i , \label{eq:Gauss}
\end{equation}
where $q_i = e b_i^* b_i$ is the charge on site $i$.  The
corresponding equation for $\kappa_{\ell}$ is the Amp\`ere-Maxwell
equation
\begin{equation}
   \sum_{d_q \in S_{\ell}} B_q d_q - \frac{1}{c}
   \frac{\text{d}}{\text{d}t} (E_{\ell} S_{\ell}) = \frac{4 \pi}{c}
   I_{\ell} , \label{eq:AmpMax}
\end{equation}
where $I_{\ell} = I_{ji} = (2 e / \hbar) \text{Im} (b_{j}^*
\tilde{H}_{ji} b_{i})$ is the current from site $i$ to site $j$.
Summing (\ref{eq:AmpMax}) over all links that contain a given site $i$
yields the charge conservation law \cite{note:charge}
\begin{equation}
   \dot{q}_i + \sum_{j} I_{ji} = 0 . \label{eq:charge}
\end{equation}
Thus, we see that $\lambda_i$ and $\kappa_{\ell}$ can be given a
consistent interpretation as discrete electromagnetic potentials,
since the above equations are in full agreement with macroscopic
(i.e., long-wavelength) electromagnetism.

In some applications of Voronoi polyhedra, it is useful to link sites
$i$ and $j$ whose polyhedra share only a line or point, hence
$S_{\ell} = S_{ij} = 0$. \cite{Okabe00} For such links, the link
volume $\Omega_{\ell} = \frac13 S_{\ell} d_{\ell}$ is zero, so the
electric field $E_{\ell}$ does not contribute to the field Lagrangian
(\ref{eq:Lf}), Gauss's law (\ref{eq:Gauss}), or the Amp\`ere-Maxwell
equation (\ref{eq:AmpMax}).  The magnetic-field contribution to
(\ref{eq:AmpMax}) likewise vanishes, because $S_{\ell} = 0$.  The
current through such a link must therefore be zero, which can only be
true in general if the Hamiltonian matrix element $H_{ij}$ vanishes.
Links with $S_{ij} = 0$ are consequently devoid of any physical
significance, and there is no loss of generality if one excludes them
at the outset by linking only sites with $S_{ij} > 0$.

Returning to the Lagrangian $L$, the Euler-Lagrange equation for
$b_i^*$ is just the Schr\"odinger equation
\begin{equation}
   i \hbar \dot{b}_i = \sum_{j} \tilde{H}_{ij} b_{j}
   . \label{eq:Schr}
\end{equation}
Since $b_i$ is an ordinary probability amplitude with a well-defined
phase, $\tilde{H}_{ij}$ must be the Hamiltonian in the presence of
electromagnetic fields.  With the restriction (\ref{eq:Hij0}), one can
express (\ref{eq:Htilde}) as
\begin{eqnarray}
   \tilde{H}_{ij} & = & H_{ij} \exp(-i \kappa_{ij}) + \hbar \lambda_i
   \delta_{ij} \nonumber \\ & = & H_{ij} \exp ( i e A_{ij} d_{ij} /
   \hbar c) + e \phi_i \delta_{ij} . \label{eq:Hfield}
\end{eqnarray}
Note the strong similarity between this result and the Peierls
substitution (\ref{eq:Peierls}).  The main difference is that
(\ref{eq:Hfield}) gives the Hamiltonian in the coordinate
representation, not the tight-binding representation.

If $|\kappa_{ij}| \ll 1$ (i.e., if the field is weak or the lattice
spacing is small), the Hamiltonian (\ref{eq:Hfield}) reduces to
\begin{equation}
   \tilde{H}_{ij} \simeq H_{ij} - \frac{e}{mc} \mathbf{A}_{ij} \cdot
   \mathbf{p}_{ij} + \frac{e^2 A_{ij}^2}{2mc^2} \Delta_{ij} + e \phi_i
   \delta_{ij} . \label{eq:H_taylor}
\end{equation}
Here a vector potential has been defined as $\mathbf{A}_{ij} = A_{ij}
\hat{\mathbf{d}}_{ij}$, while the momentum operator is given by
\begin{equation}
   \mathbf{p}_{ij} = \frac{m}{i\hbar} \mathbf{d}_{ij} H_{ij} =
   \frac{m}{i\hbar} (\mathbf{x}_{i} - \mathbf{x}_{j}) H_{ij} ,
   \label{eq:pij2}
\end{equation}
which is the same as the kinematic momentum defined above in Eq.\
(\ref{eq:pxH}).  The Hamiltonian (\ref{eq:Hfield}) therefore clearly
gives the correct first-order $\mathbf{A} \cdot \mathbf{p}$ coupling.
We shall see below that the dimensionless quantity
\begin{equation}
   \Delta_{ij} = \frac{1}{i\hbar} \mathbf{d}_{ij} \cdot
   \mathbf{p}_{ij} = - \frac{m}{\hbar^2} d_{ij}^2 H_{ij}
   \label{eq:Delta}
\end{equation}
can be viewed as a geometric weight factor that gives the correct
$A^2$ coupling also.

\subsection{Geometric definition of momentum and~kinetic~energy}

\label{sec:kinetic}

Up to this point, little has been said about the structure of the
Hamiltonian $H_{ij}$.  Within the bounds of the restriction
(\ref{eq:Hij0}), $H_{ij}$ may be treated as an arbitrary fitting
parameter.  However, in some circumstances it may be desirable to
reduce the number of fitting parameters by using a theoretical formula
for $H_{ij}$ that would reproduce the Schr\"odinger equation in the
limit of zero lattice spacing.

Let us start by considering the momentum operator, which will be
defined in this section as the canonical momentum $\mathbf{p} = -i
\hbar \nabla$.  A discrete expression for the gradient operator may be
obtained from the integral definition of the gradient: \cite{Arf85}
\begin{equation}
   \nabla f (\mathbf{x}) = \lim_{\Omega \rightarrow 0} \left[
   \frac{1}{\Omega} \int_{\partial \Omega} f(\mathbf{x}) \text{d}
   \mathbf{S} \right] ,
\end{equation}
where $\text{d} \mathbf{S}$ is a surface element pointing outward from
$\Omega$.  Now the limiting volume in a discrete lattice is the volume
$\Omega_i$ of the Voronoi polyhedron for site $\mathbf{x}_{i}$.  On
the surface $S_{ij}$ shared by $\Omega_i$ and $\Omega_{j}$, the value
of $f(\mathbf{x})$ may be taken to be $\frac{1}{2} [f(\mathbf{x}_{i})
+ f(\mathbf{x}_{j})]$.  Hence, the discrete gradient may be defined as
\begin{equation}
   \nabla f (\mathbf{x}_i) = \frac{1}{\Omega_i} \sum_{j} \frac{1}{2}
   [f(\mathbf{x}_{i}) + f(\mathbf{x}_{j})] \mathbf{S}_{ji} ,
\end{equation}
where $\mathbf{S}_{ji} = S_{ji} \hat{\mathbf{d}}_{ji}$.  Now since
\begin{equation}
   \sum_{j} \mathbf{S}_{ji} = \int_{\partial \Omega_i} \text{d}
   \mathbf{S} = 0 ,
\end{equation}
the term involving $f(\mathbf{x}_i)$ drops out, leaving only
\begin{equation}
   \nabla f (\mathbf{x}_i) = \frac{1}{2\Omega_i} \sum_{j}
   f(\mathbf{x}_{j}) \mathbf{S}_{ji} .
\end{equation}
An alternative derivation of this result is given in Eq.~(17) of
Ref.\ \onlinecite{ChrFriLee82c}.

The canonical momentum operator $\mathbf{p}$ may therefore be defined
as \cite{Sak94}
\begin{subequations}
\label{eq:p_def}
\begin{eqnarray}
   \langle \mathbf{X}_i | \mathbf{p} | \varphi \rangle & = & -i \hbar
   \nabla \langle \mathbf{X}_i | \varphi \rangle \label{eq:p_def_a} \\
   & = & - \frac{i\hbar}{2\Omega_i} \sum_{j} \langle \mathbf{X}_{j} |
   \varphi \rangle \mathbf{S}_{ji} . \label{eq:p_def_b}
\end{eqnarray}
\end{subequations}
Here the basis kets $| \mathbf{X}_i \rangle = \Omega_{i}^{-1/2} |
\mathbf{x}_{i} \rangle$ are chosen to satisfy ``$\delta$-function''
normalization
\begin{equation}
   \langle \mathbf{X}_i | \mathbf{X}_{j} \rangle =
   \frac{\delta_{ij}}{\Omega_i} , \label{eq:delta_norm}
\end{equation}
in contrast to the usual kets $| \mathbf{x}_i \rangle$, which are
normalized to unity [see Eq.\ (\ref{eq:deltaij})].  The normalization
(\ref{eq:delta_norm}) is used here because it agrees (in the limit
$\Omega_i \rightarrow 0$) with the $\delta$-function normalization of
continuous coordinate eigenkets, upon which the definition
(\ref{eq:p_def_a}) is based. \cite{Sak94}

Substituting $| \varphi \rangle = | \mathbf{X}_{j} \rangle$ in Eq.\
(\ref{eq:p_def_b}) then gives
\begin{equation}
   \langle \mathbf{X}_i | \mathbf{p} | \mathbf{X}_{j} \rangle =
   \frac{i\hbar \mathbf{S}_{ij}}{2\Omega_{i}\Omega_{j}} ,
   \label{eq:pij0}
\end{equation}
which clearly satisfies
\begin{equation}
   \sum_{j} \langle \mathbf{X}_i | \mathbf{p} | \mathbf{X}_{j} \rangle
   \Omega_{j} = 0 . \label{eq:psum}
\end{equation}
Replacing $| \mathbf{X}_i \rangle = \Omega_{i}^{-1/2} | \mathbf{x}_{i}
\rangle$ in (\ref{eq:pij0}) then yields the desired result
\begin{equation}
   \langle \mathbf{x}_i | \mathbf{p} | \mathbf{x}_{j} \rangle =
   \frac{i\hbar \mathbf{S}_{ij}}{2\sqrt{\Omega_{i}\Omega_{j}}}
   . \label{eq:pij}
\end{equation}
Note that this matrix is Hermitian, because $\mathbf{S}_{ji} =
-\mathbf{S}_{ij}$.  If the kets $| \mathbf{x}_i \rangle$ are used in
Eq.\ (\ref{eq:p_def_a}) above, a non-Hermitian canonical momentum is
obtained.

There is some question as to whether this definition of $\mathbf{p}$
should be referred to as ``canonical,'' because it does not satisfy
the canonical commutation relations.  In a continuous coordinate basis,
the canonical momentum satisfies
\begin{equation}
   \langle \mathbf{x}' | [x^{\alpha}, p^{\beta} ] | \mathbf{x}''
   \rangle = i \hbar \delta_{\alpha \beta} \delta(\mathbf{x}' -
   \mathbf{x}'') ,
\end{equation}
where $\alpha$ and $\beta$ are Cartesian components of the given
vectors.  The corresponding equation in the discrete basis is
\begin{equation}
   \langle \mathbf{X}_i | [x^{\alpha}, p^{\beta} ] | \mathbf{X}_{j}
   \rangle = \frac{i \hbar d_{ij}^{\alpha}
   S_{ij}^{\beta}}{2\Omega_{i}\Omega_{j}} ,
\end{equation}
which obviously does not agree.  Note, however, that
\begin{eqnarray}
   \sum_{i,j} \langle \mathbf{X}_i | [x^{\alpha}, p^{\beta} ] |
   \mathbf{X}_{j} \rangle \Omega_{i} \Omega_{j} & = & \frac{i
   \hbar}{2} \sum_{i,j} d_{ij}^{\alpha} S_{ij}^{\beta} \nonumber \\ &
   = & i \hbar \delta_{\alpha \beta} \Omega , \label{eq:commute}
\end{eqnarray}
where $\Omega$ is the total volume, and the second equality is proved
in Eq.\ (11) of Ref.\ \onlinecite{ChrFriLee82c}.  This agrees with the
relation
\begin{equation}
   \int\!\!\!\int \langle \mathbf{x}' | [x^{\alpha}, p^{\beta} ] |
   \mathbf{x}'' \rangle \text{d}^3 \! x' \text{d}^3 \! x'' = i \hbar
   \delta_{\alpha \beta} \Omega
\end{equation}
in the continuous basis.  Hence, Eq.\ (\ref{eq:commute}) is as close
as one can come to a canonical commutation relation in a general
discrete basis.  \cite{note:trace,note:commute}

A similar definition may be used for the kinetic energy operator $T =
- \hbar^2 \nabla^2 / 2m$.  The integral definition of the divergence,
\cite{Arf85}
\begin{equation}
   \nabla \cdot \mathbf{F} (\mathbf{x}) = \lim_{\Omega \rightarrow 0}
   \left[ \frac{1}{\Omega} \int_{\partial \Omega}
   \mathbf{F}(\mathbf{x}) \cdot \text{d} \mathbf{S} \right] ,
\end{equation}
gives the Laplacian
\begin{equation}
   \nabla^2 f(\mathbf{x}) = \lim_{\Omega \rightarrow 0} \left[
   \frac{1}{\Omega} \int_{\partial \Omega} \nabla f(\mathbf{x}) \cdot
   \text{d} \mathbf{S} \right] ,
\end{equation}
the discrete form of which is
\begin{equation}
   \nabla^2 f(\mathbf{x}_i) = \frac{1}{\Omega_i} \sum_{j} \left(
   \frac{f(\mathbf{x}_{j}) - f(\mathbf{x}_{i})}{d_{ji}} \right) S_{ji}
   .
\end{equation}
An alternative derivation of this result is given in Eq.~(12) of
Ref.\ \onlinecite{ChrFriLee82c}.  The procedure used above in Eqs.\
(\ref{eq:p_def})--(\ref{eq:pij}) then yields the kinetic energy
operator
\begin{equation}
   \langle \mathbf{x}_{i} | T | \mathbf{x}_{j} \rangle = -
   \frac{\hbar^2}{2m} \, \frac{S_{ij}}{d_{ij}
   \sqrt{\Omega_{i}\Omega_{j}}} + \delta_{ij} \,
   \frac{\hbar^2}{2m\Omega_i} \sum_{k} \frac{S_{ik}}{d_{ik}} ,
   \label{eq:Tij}
\end{equation}
which satisfies [cf.\ Eq.\ (\ref{eq:psum})]
\begin{equation}
   \sum_{j} \langle \mathbf{X}_i | T | \mathbf{X}_{j} \rangle
   \Omega_{j} = 0 . \label{eq:Tsum}
\end{equation}
Note that for $i \ne j$, $\langle \mathbf{x}_{i} | T | \mathbf{x}_{j}
\rangle$ decreases continuously to zero when $S_{ij} \rightarrow 0$.
This ensures that the Hamiltonian is a continuous function of the
lattice coordinates, even as new links are formed and old ones are
broken.

Such continuity is also desirable when the Hamiltonian is determined
empirically, especially for applications (such as molecular dynamics
\cite{WangHo96,Xu92}) in which the atomic positions vary with time.
This can be achieved by defining the nonlocal elements of the
empirical Hamiltonian as
\begin{equation}
   \langle \mathbf{x}_{i} | H | \mathbf{x}_{j} \rangle = -
   \frac{\hbar^2}{2m} \, \frac{S_{ij}}{d_{ij}
   \sqrt{\Omega_{i}\Omega_{j}}} f_{ij} \qquad (i \ne j) ,
   \label{eq:H_cont}
\end{equation}
where the fitting parameter $f_{ij}$ is a continuous, nonsingular
function of the lattice coordinates.

Note that the operators $\mathbf{p}$ and $T$ do not satisfy $T =
\mathbf{p}^2 / 2m$, because $\mathbf{p}^2$, unlike $\mathbf{p}$ and
$T$, couples sites that are not linked.  However, $\mathbf{p}$ and $T$
are related by
\begin{subequations}
\label{eq:pT}
\begin{equation}
   \mathbf{p}_{ij} = \frac{m}{i\hbar} \mathbf{d}_{ij} T_{ij}
   \label{eq:pT1}
\end{equation}
or
\begin{equation}
   \mathbf{p} = \frac{m}{i\hbar} [ \mathbf{x}, T ] . \label{eq:pT2}
\end{equation}
\end{subequations}
Thus, any Hamiltonian of the form $H = T + V(\mathbf{x})$, where $V$
is a local potential, satisfies Eq.\ (\ref{eq:pxH}).  Hence, for such
a Hamiltonian, the canonical momentum $\mathbf{p} = -i \hbar \nabla$
used in this section agrees with the kinematic momentum defined
earlier.

Now let us examine the dimensionless factor $\Delta_{ij}$ defined
above in Eq. (\ref{eq:Delta}).  If $H = T + V$, this becomes
\begin{equation}
   \Delta_{ij} = \frac{S_{ij} d_{ij}}{2 \sqrt{\Omega_{i}\Omega_{j}}} =
   \frac{3 \Omega_{ij}}{2 \sqrt{\Omega_{i}\Omega_{j}}} ,
\end{equation}
where $\Omega_{ij} = \frac13 S_{ij} d_{ij}$ is the volume of the link
between sites $i$ and $j$ (see Appendix \ref{app:geometry}).  The
factor $\Delta_{ij}$ appears in the $A^2$ term in the Hamiltonian
(\ref{eq:H_taylor}), which will be referred to as $H_2$.  In a
continuous coordinate basis, $H_2$ is given by
\begin{equation}
   \langle \mathbf{x}' | H_2 | \mathbf{x}'' \rangle =
   \frac{e^2}{2mc^2} A^2(\mathbf{x}') \delta(\mathbf{x}' -
   \mathbf{x}'') ,
\end{equation}
which means that it satisfies
\begin{equation}
   \int \!\!\! \int \langle \mathbf{x}' | H_2 | \mathbf{x}'' \rangle
   \text{d}^3\! x' \text{d}^3 \! x'' = \frac{e^2}{2mc^2} \int
   A^2(\mathbf{x}') \text{d}^3 \! x' . \label{eq:A2int}
\end{equation}
The corresponding equations for the discrete basis are
\begin{equation}
   \langle \mathbf{X}_i | H_2 | \mathbf{X}_j \rangle =
   \frac{e^2}{2mc^2} \frac{3 A_{ij}^2 \Omega_{ij}}{2 \Omega_{i}
   \Omega_{j}}
\end{equation}
and
\begin{eqnarray}
   \sum_{i,j} \langle \mathbf{X}_i | H_2 | \mathbf{X}_j \rangle
   \Omega_i \Omega_j & = & \frac{e^2}{2mc^2} \frac32 \sum_{i,j}
   A_{ij}^2 \Omega_{ij} \nonumber \\ & = & \frac{e^2}{2mc^2}
   \sum_{\ell} 3 A_{\ell}^2 \Omega_{\ell} . \label{eq:A2sum}
\end{eqnarray}
The second equality in (\ref{eq:A2sum}) was obtained by noting that a
sum over $i$ and $j$ covers each link $\ell$ twice.  The only apparent
difference between Eqs.\ (\ref{eq:A2int}) and (\ref{eq:A2sum}) is a
factor of 3.  This appears for the same reason that it does in the
Lagrangian (\ref{eq:Lf})---i.e., $A^2$ in Eq.\ (\ref{eq:A2int}) refers
to $A_x^2 + A_y^2 + A_z^2$, whereas $A_{\ell}^2$ in Eq.\
(\ref{eq:A2sum}) refers only to the component of $\mathbf{A}$ in the
direction of $\mathbf{d}_{\ell}$.

Therefore, Eqs.\ (\ref{eq:A2int}) and (\ref{eq:A2sum}) are the same in
the limit of zero lattice spacing, and the factor $\Delta_{ij}$ is
simply a geometric weight factor that provides the correct $A^2$
coupling in the Hamiltonian (\ref{eq:H_taylor}).

\subsection{Spin}

\label{sec:spin}

The theory presented thus far has been for a particle with spin zero.
Particles with spin $\frac12$ may be described using a discrete
version of the Dirac Hamiltonian for a free particle:
\begin{equation}
   H = c \bm{\alpha} \cdot \mathbf{p} + \beta m c^2 ,
\end{equation}
where $\bm{\alpha}$ and $\beta$ are Dirac's $4 \times 4$ matrices.
The momentum operator $\mathbf{p}$ can either be calculated from
geometry or fitted to experiment.  In the presence of electromagnetic
fields, the Hamiltonian becomes
\begin{equation}
   \tilde{H} = c \bm{\alpha} \cdot \bm{\pi} + \beta m c^2 + e \phi ,
   \label{eq:HDirac}
\end{equation}
where [cf.\ Eq.\ (\ref{eq:Hfield})]
\begin{equation}
   \bm{\pi}_{ij} = \mathbf{p}_{ij} \exp (-i \kappa_{ij}) .
\end{equation}
A nonrelativistic Hamiltonian may be obtained by applying a
Foldy-Wouthuysen transformation \cite{FolWou50,Mess62} to Eq.\
(\ref{eq:HDirac}), which yields
\begin{eqnarray}
   H_{\text{nr}} & = & \frac{1}{2m} (\bm{\sigma} \cdot \bm{\pi})^2 -
   \frac{1}{8m^3c^2} (\bm{\sigma} \cdot \bm{\pi})^4 + e \phi
   \label{eq:Hnr} \\ && {} - \frac{1}{8m^2c^2} [ \bm{\sigma} \cdot
   \bm{\pi} , ( [ \bm{\sigma} \cdot \bm{\pi} , e \phi ] + i \hbar
   \bm{\sigma} \cdot \dot{\bm{\pi}} ) ] , \nonumber
\end{eqnarray}
where $\bm{\sigma}$ is the Pauli spin matrix, and all terms of order
$(v/c)^4$ have been included.  This Hamiltonian couples sites that are
not linked, but there is no ambiguity because the Dirac equation is
taken as fundamental.

If we assume for simplicity that the lattice coordinates do not depend
on time, then
\begin{equation}
   ( [ \bm{\sigma} \cdot \bm{\pi} , e \phi ] + i \hbar \bm{\sigma}
   \cdot \dot{\bm{\pi}} )_{ij} = - \bm{\sigma} \cdot \bm{\pi}_{ij}
   V_{ij} ,
\end{equation}
where
\begin{equation}
   V_{ij} = e (\phi_i - \phi_j) - \hbar \dot{\kappa}_{ij} = - e E_{ij}
   d_{ij} 
\end{equation}
is the difference in potential energy of sites $i$ and $j$ due to the
electric field.  The last term in Eq.\ (\ref{eq:Hnr}) therefore
consists of the Darwin term
\begin{equation}
   H_{ij}^{\text{D}} = \frac{1}{8m^2c^2} \sum_{k} (V_{ki} + V_{kj})
   \bm{\pi}_{ik} \cdot \bm{\pi}_{kj} 
\end{equation}
plus the spin-orbit coupling
\begin{equation}
   H_{ij}^{\text{so}} = \frac{i}{8m^2c^2} \sum_{k} (V_{ki} + V_{kj})
   \bm{\sigma} \cdot (\bm{\pi}_{ik} \times \bm{\pi}_{kj}) ,
   \label{eq:Hso}
\end{equation}
where the identity
\begin{equation}
   (\bm{\sigma} \cdot \mathbf{a}) (\bm{\sigma} \cdot \mathbf{b}) =
   \mathbf{a} \cdot \mathbf{b} + i \bm{\sigma} \cdot (\mathbf{a}
   \times \mathbf{b})
\end{equation}
has been used.  Now the main contribution to spin-orbit coupling comes
from the atomic cores, where the potential energy and wave function
vary rapidly.  However, in any basis of reasonable size, the lattice
imposes a wavelength cutoff that eliminates such rapid variations.
The potential $\phi_i$ must therefore be viewed as a pseudopotential,
not a true atomic potential.  Hence, for practical purposes, $\phi_i$
in the spin-orbit Hamiltonian (\ref{eq:Hso}) should be treated as a
fitting parameter that is independent of the value used for the first
$e \phi$ term in (\ref{eq:Hnr}).

The first two terms in the Hamiltonian (\ref{eq:Hnr}) are
kinetic-energy terms, which may be rewritten using
\begin{equation}
   (\bm{\sigma} \cdot \bm{\pi})_{ij}^2 = \sum_{k} [ \bm{\pi}_{ik}
   \cdot \bm{\pi}_{kj} + i \bm{\sigma} \cdot (\bm{\pi}_{ik} \times
   \bm{\pi}_{kj}) ] ,
\end{equation}
in which the second term describes the intrinsic magnetic dipole
moment of the particle.  For a general lattice, this term is not zero
even when there is no electromagnetic field, because different
components of the momentum operator do not commute (i.e., $\mathbf{p}
\times \mathbf{p} \ne 0$).  This follows from the fact that there is
generally no more than one site $k$ linked to both $i$ and $j$, and
for that $i$ and $j$, $\mathbf{p}_{ik} \times \mathbf{p}_{kj}$ is
generally not zero.

However, if the lattice is a Bravais lattice, then $\mathbf{p} \times
\mathbf{p}$ is always zero.  This follows from the fact that every
site in a Bravais lattice is identical, so for a given nonzero
$\mathbf{p}_{ik} \times \mathbf{p}_{kj}$, there is always another site
$l$ with $\mathbf{p}_{il} = \mathbf{p}_{kj}$ and $\mathbf{p}_{lj} =
\mathbf{p}_{ik}$, hence $\mathbf{p}_{ik} \times \mathbf{p}_{kj} +
\mathbf{p}_{il} \times \mathbf{p}_{lj} = 0$.  Clearly one also has
$\mathbf{d}_{ik} \times \mathbf{d}_{kj} + \mathbf{d}_{il} \times
\mathbf{d}_{lj} = 0$, so the sites $i$, $k$, $j$, and $l$ lie in a
single plane.  If $i$ is not linked to $j$ and $k$ is not linked to
$l$, then $i$, $k$, $j$, and $l$ form a single plaquette $q$, which
has the shape of a rectangle.  Otherwise, they form two triangular
plaquettes.

If the momentum operator is given by Eq.\ (\ref{eq:pij}), then the
intrinsic magnetic dipole term in the Hamiltonian (\ref{eq:Hnr}) for
such a Bravais lattice is
\begin{equation}
   H^{\text{mag}}_{ij} = - \frac{e \hbar}{8mc} \left( \frac{S_{ik}
   S_{kj} S_q^2}{d_{ik} d_{kj} \Omega_i^2} \right) \bm{\sigma} \cdot
   \mathbf{B}_q , \label{eq:Hmag}
\end{equation}
where the weak-field approximation $|\kappa_{ij}| \ll 1$ has been
used, and the direction of $\mathbf{B}_q$ is that of $\mathbf{S}_q$.
If the sites $i$, $k$, $j$, and $l$ form a single rectangular
plaquette, then $S_q$ is the area of that plaquette; otherwise, it is
the combined area of the two triangles (in which case $\mathbf{B}_q$
is the average magnetic field of the two plaquettes).

As an example, consider a simple cubic lattice with lattice constant
$a$, for which $S_{ik} = S_{kj} = S_{q} = a^2$, $d_{ik} = d_{kj} = a$,
and $\Omega_i = a^3$.  In this case, the factor in parentheses in Eq.\
(\ref{eq:Hmag}) is unity, and $H^{\text{mag}}_{ij}$ couples sites on
opposite corners of each plaquette (with $d_{ij} = \sqrt{2} a$).  By
comparison, the dipole term in the continuum Hamiltonian is given by
\begin{equation}
   \frac{1}{2m} [\bm{\sigma} \cdot (\mathbf{p} - e \mathbf{A} / c)]^2
   = \frac{1}{2m} (\mathbf{p} - e \mathbf{A} / c)^2 - \frac{e
   \hbar}{2mc} \bm{\sigma} \cdot \mathbf{B} .
\end{equation}
The numerical factor in front of this dipole coupling is four times
larger than that in Eq.\ (\ref{eq:Hmag}).  This occurs because
(\ref{eq:Hmag}) couples each site $i$ to four other sites $j$.

\section{Application to tetrahedral semiconductors}

\label{sec:application}

This section considers several different methods of implementing the
theory developed in Sec.\ \ref{sec:gauge}.  Spin is neglected in all
of the applications that follow.

\subsection{Discrete pseudopotential method}

\label{subsec:pseudo}

The simplest geometry occurs when the lattice sites $\mathbf{x}_i$ are
chosen to lie on a Bravais lattice.  One possible approach in this
case is to use the geometric expression (\ref{eq:Tij}) for the kinetic
energy $T$, and assume that the potential energy $V$ is local.  This
approach will be referred to as the discrete pseudopotential method.

If $\mathbf{x}$ lies on a Bravais lattice (the subscript $i$ is
omitted here), one may define a reciprocal lattice as the set of all
vectors $\mathbf{g}$ such that $\mathbf{g} \cdot \mathbf{x} = 2 \pi
\times \text{integer}$.  The volume of a primitive cell in the direct
lattice is denoted $\omega_0$, while that of a primitive cell in the
reciprocal lattice is $\omega_0^* = (2\pi)^3 / \omega_0$.

In a crystal, the Hamiltonian will be periodic with respect to some
larger Bravais lattice whose sites are denoted by $\mathbf{R}$, where
$\mathbf{R} \in \{ \mathbf{x} \}$.  One may then define a
corresponding reciprocal lattice as the set of all vectors
$\mathbf{G}$ such that $\mathbf{G} \cdot \mathbf{R} = 2 \pi
\times \text{integer}$.  The volume of a primitive cell in $\mathbf{R}$
space is $\Omega_0 = n \omega_0$, where $n$ is an integer, and the
volume of a primitive cell in $\mathbf{G}$ space is $\Omega_0^* =
(2\pi)^3 / \Omega_0$.

Periodic boundary conditions may then be implemented over an even
larger Bravais lattice whose sites are denoted by $\mathbf{L}$, where
$\mathbf{L} \in \{ \mathbf{R} \}$.  The corresponding reciprocal
lattice vectors are denoted $\mathbf{k}$, where $\mathbf{k} \cdot
\mathbf{L} = 2 \pi \times \text{integer}$.  The volume over which
periodic boundary conditions are applied is $\Omega = N \Omega_0$,
where $N$ is an integer, and the volume of a primitive cell in
$\mathbf{k}$ space is $\Omega^* = (2\pi)^3 / \Omega$.  Note that
according to the above definitions, $\mathbf{G} \in \{ \mathbf{k} \}$
and $\mathbf{g} \in \{ \mathbf{G} \}$.

Coordinate eigenkets in this system are denoted $| \mathbf{x}
\rangle$, where $| \mathbf{x} + \mathbf{L} \rangle \equiv | \mathbf{x}
\rangle$ due to the periodic boundary conditions.  The orthogonality
and closure relations in this basis are therefore
\begin{eqnarray}
   && \langle \mathbf{x} | \mathbf{x}' \rangle = \sum_{\mathbf{L}}
   \delta_{\mathbf{x} - \mathbf{x}', \mathbf{L}} , \\ &&
   \sum_{\mathbf{x} \in \Omega} | \mathbf{x} \rangle \langle
   \mathbf{x} | = 1 .
\end{eqnarray}
In a system with periodic boundary conditions, the coordinate operator
is not well defined; only periodic functions of the coordinate are
permitted.  Therefore, the definition of the kinematic momentum must
be slightly modified.  Instead of (\ref{eq:pxH}), one has
\begin{equation}
   \langle \mathbf{x} | \mathbf{p} | \mathbf{x}' \rangle = \frac{m}{i
   \hbar} ( \mathbf{x} - \mathbf{x}' ) \langle \mathbf{x} | H |
   \mathbf{x}' \rangle \qquad \text{if } \mathbf{x} \in \Omega \text{
   and } \mathbf{x}' \in \Omega , \label{eq:p_periodic}
\end{equation}
with $\langle \mathbf{x} | \mathbf{p} | \mathbf{x}' \rangle = \langle
\mathbf{x} + \mathbf{L} | \mathbf{p} | \mathbf{x}' \rangle$ otherwise.

Another useful representation is the crystal momentum representation
\begin{equation}
   | \mathbf{k} \rangle = \frac{1}{\sqrt{\mathcal{N}}}
     \sum_{\mathbf{x} \in \Omega} e^{i \mathbf{k} \cdot \mathbf{x}} |
     \mathbf{x} \rangle ,
\end{equation}
where $\mathcal{N} = nN = \Omega / \omega_0$.  The corresponding
orthogonality and closure relations are
\begin{eqnarray}
   && \langle \mathbf{k} | \mathbf{k}' \rangle = \sum_{\mathbf{g}}
   \delta_{\mathbf{k} - \mathbf{k}', \mathbf{g}} , \\ &&
   \sum_{\mathbf{k} \in \omega_0^*} | \mathbf{k} \rangle \langle
   \mathbf{k} | = 1 .
\end{eqnarray}
In the crystal momentum representation, a periodic Hamiltonian
$\langle \mathbf{x} | H | \mathbf{x}' \rangle = \langle \mathbf{x} +
\mathbf{R} | H | \mathbf{x}' + \mathbf{R} \rangle$ couples only those
states that differ by a reciprocal lattice vector $\mathbf{G}$:
\begin{equation}
   \langle \mathbf{k}' | H | \mathbf{k} \rangle = \sum_{\mathbf{G}}
   \delta_{\mathbf{k}', \mathbf{k} + \mathbf{G}} \langle \mathbf{k} +
   \mathbf{G} | H | \mathbf{k} \rangle , \label{eq:Hk}
\end{equation}
where
\begin{equation}
   \langle \mathbf{k} + \mathbf{G} | H | \mathbf{k} \rangle =
   \frac{1}{n} \sum_{\mathbf{x} \in \Omega} \sum_{\mathbf{x}' \in
   \Omega_0} e^{-i (\mathbf{k} + \mathbf{G}) \cdot \mathbf{x}'}
   \langle \mathbf{x}' | H | \mathbf{x} \rangle e^{i \mathbf{k} \cdot
   \mathbf{x}} .
\end{equation}
The kinematic momentum (\ref{eq:p_periodic}) also satisfies Eq.\
(\ref{eq:Hk}).  Its matrix elements that are related to those of $H$
by
\begin{equation}
   \langle \mathbf{k} + \mathbf{G} | \mathbf{p} | \mathbf{k} \rangle =
   \frac{m}{\hbar} \nabla_{\mathbf{k}} \langle \mathbf{k} + \mathbf{G}
   | H | \mathbf{k} \rangle . \label{eq:pk}
\end{equation}

On a Bravais lattice, the kinetic energy (\ref{eq:Tij}) and canonical
momentum (\ref{eq:pij}) are translationally invariant:
\begin{equation}
   \langle \mathbf{x} | T | \mathbf{x}' \rangle = T(\mathbf{x} -
   \mathbf{x}') ,
\end{equation}
where $T(\mathbf{x} + \mathbf{L}) = T(\mathbf{x})$.  The matrix
elements of $T$ are therefore given by
\begin{equation}
   \langle \mathbf{k} | T | \mathbf{k}' \rangle = T(\mathbf{k})
   \sum_{\mathbf{g}} \delta_{\mathbf{k} - \mathbf{k}', \mathbf{g}} ,
\end{equation}
where
\begin{equation}
   T(\mathbf{k}) = \sum_{\mathbf{x} \in \Omega} T(\mathbf{x}) e^{-i
   \mathbf{k} \cdot \mathbf{x}} . \label{eq:Tk}
\end{equation}
For tetrahedral semiconductors with the diamond or zinc-blende
structure, it is convenient to use a cubic lattice for the grid $\{
\mathbf{x} \}$.  Expressions for the link distances $d_{ij}$ and
surface areas $S_{ij}$ are given in Appendix \ref{app:link} for the
simple cubic, body-centered cubic, and face-centered cubic lattices.
The resulting kinetic-energy operators given by Eqs.\ (\ref{eq:Tij})
and (\ref{eq:Tk}) are
\begin{eqnarray}
   T_{\text{sc}}(\mathbf{k}) & = & \frac{2 \hbar^2}{ma^2} [ \sin^2
   (\textstyle\frac12 k_x a) + \sin^2 (\textstyle\frac12 k_y a) +
   \sin^2 (\textstyle\frac12 k_z a) ] , \nonumber \\
   T_{\text{bcc}}(\mathbf{k}) & = & \frac{\hbar^2}{2ma^2} \{ 6 [ 1 -
   \cos (\textstyle\frac12 k_x a) \cos (\textstyle\frac12 k_y a) \cos
   (\textstyle\frac12 k_z a) ] \nonumber \\ && {} + [ \sin^2
   (\textstyle\frac12 k_x a) + \sin^2 (\textstyle\frac12 k_y a) +
   \sin^2 (\textstyle\frac12 k_z a) ] \} , \nonumber \\
   T_{\text{fcc}}(\mathbf{k}) & = & \frac{2 \hbar^2}{ma^2} [ 3 - \cos
   (\textstyle\frac12 k_y a) \cos (\textstyle\frac12 k_z a) \nonumber
   \\ && {} - \cos (\textstyle\frac12 k_z a) \cos (\textstyle\frac12
   k_x a) \nonumber \\ && {} - \cos (\textstyle\frac12 k_x a) \cos
   (\textstyle\frac12 k_y a) ] , \label{eq:Tfcc}
\end{eqnarray}
all of which reduce to $T(\mathbf{k}) \simeq \hbar^2 k^2 / 2m$ when
$ka \ll 1$.  Here $a$ is the lattice constant of the grid $\{
\mathbf{x} \}$, which is some integer fraction of the lattice constant
$a_0$ of the crystal lattice $\{ \mathbf{R} \}$.

A canonical momentum operator $\mathbf{p}(\mathbf{k})$ corresponding
to Eq.\ (\ref{eq:pij}) may be defined in a similar manner.  This
operator is given by
\begin{equation}
   \mathbf{p}(\mathbf{k}) = \frac{m}{\hbar} \nabla_{\mathbf{k}}
   T(\mathbf{k}) ,
\end{equation}
which follows from Eq.\ (\ref{eq:pT}).  Note that this result is just
a special case of the kinematic momentum (\ref{eq:pk}).

The matrix elements of a local periodic potential $V(\mathbf{x}) =
V(\mathbf{x} + \mathbf{R})$ are given by Eq.\ (\ref{eq:Hk}), where
\begin{equation}
   \langle \mathbf{k} + \mathbf{G} | V | \mathbf{k} \rangle =
   \frac{1}{n} \sum_{\mathbf{x} \in \Omega_0} V(\mathbf{x}) e^{-i
   \mathbf{G} \cdot \mathbf{x}}
\end{equation}
is independent of $\mathbf{k}$.  It will be assumed here that $V$ is a
superposition of local atomic pseudopotentials:
\begin{equation}
   V(\mathbf{x}) = \sum_{\mu = 1}^{N_{a}} \sum_{\mathbf{R} \in
   \Omega} v_{\mu} (\mathbf{x} - \mathbf{R} - \bm{\tau}_{\mu}) ,
\end{equation}
where $v_{\mu}(\mathbf{x}) = v_{\mu}(\mathbf{x} + \mathbf{L})$
is the pseudopotential for atom $\mu$, whose position in the unit
cell $\Omega_0$ is given by $\bm{\tau}_{\mu}$.  In this case
\begin{equation}
   \langle \mathbf{k} + \mathbf{G} | V | \mathbf{k} \rangle =
   \frac{1}{N_{a}} \sum_{\mu = 1}^{N_{a}} v_{\mu}
   (\mathbf{G}) e^{-i \mathbf{G} \cdot \bm{\tau}_{\mu}} ,
\end{equation}
where $v_{\mu} (\mathbf{G})$ is the atomic form factor
\begin{equation}
   v_{\mu} (\mathbf{G}) = \frac{N_{a}}{n} \sum_{\mathbf{x} \in
   \Omega} v_{\mu} (\mathbf{x}) e^{-i \mathbf{G} \cdot \mathbf{x}}
   , \label{eq:vff}
\end{equation}
and $N_{a}$ is the number of atoms in the unit cell $\Omega_0$.

The main practical difficulty in implementing the discrete
pseudopotential method is that $T(\mathbf{k})$ is not a good
approximation to the continuum kinetic energy
\begin{equation}
   T_{\text{cont}} (\mathbf{k}) = \frac{\hbar^2 k^2}{2m}
   \label{eq:Tcont}
\end{equation}
unless $ka \ll 1$.  This means that the lattice constant $a$ of the
grid $\{ \mathbf{x} \}$ must be significantly smaller than the lattice
constant $a_0$ of the crystal lattice $\{ \mathbf{R} \}$.  To obtain
one grid point at each atom in the diamond structure, $a$ must satisfy
$a = a_0 / 2l$ (for a bcc grid) or $a = a_0 / 4l$ (for sc and fcc),
where $l$ is a positive integer.  Numerical accuracy generally
requires $l > 1$, as shown below.

The {\em shape} of the Brillouin zone is also important.  If the shape
of the Wigner-Seitz cell for $\omega_0^*$ is not congruent with the
shape of the Wigner-Seitz cell for $\Omega_0^*$, then $T(\mathbf{k})$
will deviate from $T_{\text{cont}}(\mathbf{k})$ more rapidly in some
directions than others.  This can lead to significant qualitative
errors in the kinetic energy.  For example, in diamond, the lowest
continuum eigenvalues at $\Gamma$ are $T_{\text{cont}}(\mathbf{G}) =
\hbar^2 G^2 / 2m$, where $\mathbf{G} = (000)$, $\langle 111 \rangle$,
$\langle 200 \rangle$, and $\langle 220 \rangle$ (in units of $2 \pi /
a_0$).  But for a sc grid with $a = \frac14 a_0$, the value of
$T_{\text{sc}}(\langle 200 \rangle) = 2 \hbar^2 / m a^2$ is actually
{\em lower} than that of $T_{\text{sc}}(\langle 111 \rangle) = 3
\hbar^2 / m a^2$.  The correct ratio $T_{\text{cont}}(\langle 200
\rangle) = \frac43 T_{\text{cont}}(\langle 111 \rangle)$ is only
approached in the limit of very small $a/a_0$, making the sc grid a
poor choice for diamond.

The natural choice for diamond is the fcc grid, since its Brillouin
zone has the same shape as that of diamond.  Indeed, one has
$T_{\text{fcc}}(\langle 200 \rangle) = \frac43 T_{\text{fcc}}(\langle
111 \rangle)$ even for the maximum grid size $a = \frac14 a_0$.  The
only problem here is that $T_{\text{fcc}}(\langle 220 \rangle) =
\frac32 T_{\text{fcc}}(\langle 200 \rangle)$, which is not
sufficiently close to the correct ratio $T_{\text{cont}}(\langle 220
\rangle) = 2 T_{\text{cont}}(\langle 200 \rangle)$ to make $a =
\frac14 a_0$ a satisfactory choice for numerical calculations.  The
next possibility is $a = \frac18 a_0$, which yields
$T_{\text{fcc}}(\langle 220 \rangle) = (\frac32 + \frac{\sqrt{2}}{4})
T_{\text{fcc}}(\langle 200 \rangle) \simeq 1.85 T_{\text{fcc}}(\langle
200 \rangle)$.

\begin{figure}
\includegraphics[width=3.375in,keepaspectratio]{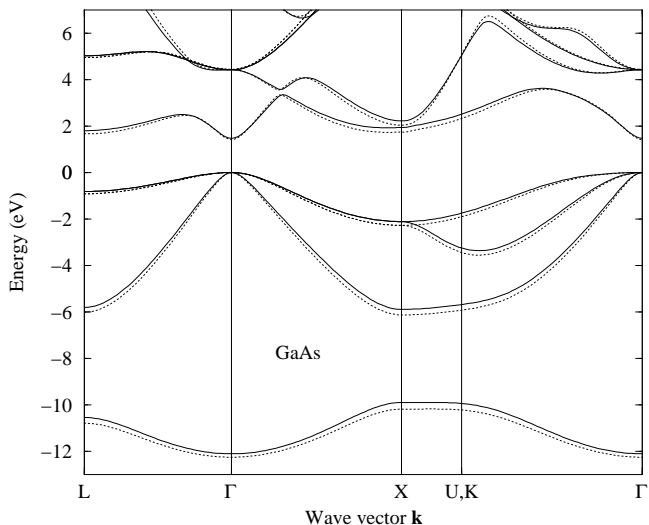}
\caption{\label{fig:GaAs} Energy band structure of GaAs calculated by
the discrete local pseudopotential method using an fcc grid with $a =
\frac18 a_0$.  Solid lines: fcc kinetic energy from Eq.\
(\ref{eq:Tfcc}).  Dotted lines: continuum kinetic energy from Eq.\
(\ref{eq:Tcont}).}
\end{figure}

The energy band structure for GaAs calculated using an fcc grid with
$a = \frac18 a_0$ is given in Fig.~\ref{fig:GaAs}.
The fcc kinetic energy obtained from Eq.\ (\ref{eq:Tfcc}) was
multiplied by a constant factor $\pi^2(2+\sqrt{2})/32 = 1.053$ so that
$T_{\text{fcc}}(\mathbf{k})$ matches $T_{\text{cont}}(\mathbf{k})$ at
$\mathbf{G} = \langle 111 \rangle$ and $\mathbf{G} = \langle 200
\rangle$.  The pseudopotential form factors (\ref{eq:vff}) for this
calculation were taken from Ref.\ \onlinecite{CoBe66}.  No attempt was
made to fit the energy bands by modifying the form factors; the
purpose of this figure is merely to demonstrate the close similarity
between the discrete fcc band structure and the continuum band
structure.  Slight adjustments in the model parameters would likely
give an even better agreement.

The main problem with this result is that $a = \frac18 a_0$
corresponds to a basis size of 512 grid points per primitive unit cell
$\Omega_0$.  This is unattractive in comparison to the basis
dimensions of approximately 100 plane waves that are typically used in
empirical pseudopotential calculations for tetrahedral semiconductors.
However, changing the fcc grid to $a = \frac14 a_0$ (i.e., 64 grid
points per primitive cell) makes it impossible to achieve a
satisfactory fit to the band structure using local pseudopotentials.
A good fit is possible only if the nonlocal Hamiltonian matrix
elements are treated as fitting parameters.  But in that case, one can
reduce the basis dimensions even further by using the tight-binding
approach.

\subsection{Tight-binding method}

In the tight-binding approach, the grid points are no longer
restricted to lie on a Bravais lattice, and all of the Hamiltonian
matrix elements are treated as fitting parameters.  In addition, it is
assumed here that the model is constructed using the symmetrization
procedure described in Sec.\ \ref{sec:coordinate}, so that a distinct
set of orthogonal orbitals is associated with each atom.  The
objective then is to find the smallest coordinate basis that provides
a physically reasonable model of a given system.

The basis kets in the tight-binding approach will be written as $|
\alpha, \mathbf{R} + \bm{\tau}_{\mu} \rangle$, where $\mathbf{R}$ is a
lattice vector for the Bravais lattice over which the Hamiltonian is
periodic, and $\bm{\tau}_{\mu}$ is the position of atom $\mu$ within
the primitive unit cell $\Omega_0$.  These quantities are defined
exactly as they were in Sec.\ \ref{subsec:pseudo}; the vectors
$\mathbf{L}$, $\mathbf{G}$, and $\mathbf{k}$ are also defined in the
same way.  The label $\alpha$ may refer to a coordinate
$\mathbf{x}_{\alpha}$ within the atom, in which case
\begin{equation}
   | \alpha, \mathbf{R} + \bm{\tau}_{\mu} \rangle \equiv |
   \mathbf{x}_{\alpha} + \mathbf{R} + \bm{\tau}_{\mu} \rangle
   \label{eq:tb_basis}
\end{equation}
is just a coordinate eigenket.  However, $\alpha$ may also be used as
a symmetry label for an atomic orbital that is a symmetrized linear
combination of the kets (\ref{eq:tb_basis}):
\begin{equation}
   | \alpha, \mathbf{R} + \bm{\tau}_{\mu} \rangle = \sum_{\beta}
   C_{\beta} (\alpha) | \mathbf{x}_{\beta} + \mathbf{R} +
   \bm{\tau}_{\mu} \rangle . \label{eq:tb_basis2}
\end{equation}
In either case, the basis is orthogonal:
\begin{equation}
   \langle \alpha, \mathbf{R} + \bm{\tau}_{\mu} | \alpha', \mathbf{R}'
   + \bm{\tau}_{\mu'} \rangle = \delta_{\alpha\alpha'}
   \delta_{\mu\mu'} \sum_{\mathbf{L}} \delta_{\mathbf{R} -
   \mathbf{R}', \mathbf{L}} ,
\end{equation}
and complete:
\begin{equation}
   \sum_{\alpha} \sum_{\mu} \sum_{\mathbf{R} \in \Omega} | \alpha,
   \mathbf{R} + \bm{\tau}_{\mu} \rangle \langle \alpha, \mathbf{R} +
   \bm{\tau}_{\mu} | = 1 .
\end{equation}

In periodic systems, it is convenient to define the Bloch sums
\cite{Bloch28}
\begin{equation}
   | \alpha, \mu, \mathbf{k} \rangle = \frac{1}{\sqrt{N}}
   \sum_{\mathbf{R} \in \Omega} e^{i \mathbf{k} \cdot (\mathbf{R} +
   \bm{\tau}_{\mu})} | \alpha, \mathbf{R} + \bm{\tau}_{\mu} \rangle ,
   \label{eq:Bloch_sum}
\end{equation}
which are also orthogonal and complete:
\begin{equation}
   \langle \alpha, \mu, \mathbf{k} | \alpha', \mu', \mathbf{k}'
   \rangle = \delta_{\alpha\alpha'} \delta_{\mu\mu'} \sum_{\mathbf{G}}
   \delta_{\mathbf{k} - \mathbf{k}', \mathbf{G}} \, e^{-i \mathbf{G}
   \cdot \bm{\tau}_{\mu}} ,
\end{equation}
\begin{equation}
   \sum_{\alpha} \sum_{\mu} \sum_{\mathbf{k} \in \Omega_0^*} | \alpha,
   \mu, \mathbf{k} \rangle \langle \alpha, \mu, \mathbf{k} | = 1 .
\end{equation}
If the Hamiltonian $H$ is invariant with respect to lattice
translations $\mathbf{R}$, then its matrix elements in the Bloch basis
are
\begin{eqnarray}
   \langle \alpha, \mu, \mathbf{k} | H | \alpha', \mu',
   \mathbf{k}' \rangle & = & \langle \alpha, \mu, \mathbf{k} | H |
   \alpha', \mu', \mathbf{k} \rangle \\ && \times \sum_{\mathbf{G}}
   \delta_{\mathbf{k} - \mathbf{k}', \mathbf{G}} e^{-i \mathbf{G}
   \cdot \bm{\tau}_{\mu'}} , \nonumber
\end{eqnarray}
where
\begin{eqnarray}
   \langle \alpha, \mu, \mathbf{k} | H | \alpha', \mu', \mathbf{k}
   \rangle & = & \sum_{\mathbf{R}' \in \Omega} \langle \alpha,
   \bm{\tau}_{\mu} | H | \alpha', \mathbf{R}' + \bm{\tau}_{\mu'}
   \rangle \nonumber \\ && {} \times e^{i \mathbf{k} \cdot
   (\mathbf{R}' + \bm{\tau}_{\mu'} - \bm{\tau}_{\mu})} . \label{eq:H_Bloch}
\end{eqnarray}
The kinematic momentum operator (\ref{eq:pij2}) is related to this
Hamiltonian by
\begin{eqnarray}
   \langle \alpha, \mu, \mathbf{k} | \mathbf{p} | \alpha', \mu',
   \mathbf{k} \rangle & = & \frac{m}{i \hbar} (\mathbf{x}_{\alpha} -
   \mathbf{x}_{\alpha'} + i \nabla_{\mathbf{k}} ) \nonumber \\ &&
   \times \langle \alpha, \mu, \mathbf{k} | H | \alpha', \mu',
   \mathbf{k} \rangle , \label{eq:pH_tb}
\end{eqnarray}
where the original basis was assumed to be given by
(\ref{eq:tb_basis}).  Note that (\ref{eq:pH_tb}) has the form of an
intra-atomic matrix element (proportional to $\mathbf{x}_{\alpha} -
\mathbf{x}_{\alpha'}$) plus an inter-atomic matrix element
(proportional to $i \nabla_{\mathbf{k}}$).  The zero-parameter model
of Eq.\ (\ref{eq:x0}) is obtained in the limit $\mathbf{x}_{\alpha}
\rightarrow 0$.

Let us now consider specific examples of $H$ for tetrahedral
semiconductors.  The simplest tight-binding model for the diamond or
zinc-blende structure consists of a single $s$ orbital per atom, which
is obtained by putting one coordinate eigenket at each atomic position
[see Eq.\ (\ref{eq:symm_basis})].  In this model, as shown in Appendix
\ref{app:link}, each atom is linked to 4 nearest neighbors and 12
second-nearest neighbors.  More distant linkages do not exist because
the Voronoi polyhedra for these atoms do not touch one another.

Such a simple model is, of course, unable to describe even the
qualitative features of tetrahedral semiconductors.  The simplest
conventional tight-binding model that works in this case is the $sp^3$
model.  \cite{SlaKos54,Har89,Har99,Tser93} The basic features of this
model are described in Table~\ref{table:freeparam}, which lists the
basis size (number of orbitals per atom) and the number of independent
Hamiltonian matrix elements for coupling between atoms in the diamond
structure out to third nearest neighbors. \cite{Tser93}
\begin{table}
\caption{\label{table:freeparam}Number of free parameters for
different tight-binding models in the diamond structure.  The upper
half of the table refers to conventional tight-binding models, while
the lower half refers to models obtained from symmetrized coordinate
eigenkets.}
\begin{ruledtabular}
\begin{tabular}{lc|cccc}
\multicolumn{2}{c|}{Model} & \multicolumn{4}{c}{Parameters} \\
Basis & Size & On-site & 1NN & 2NN & 3NN \\ \hline
$sp^3$                & 4 & 2 &  4 &  7 &  7 \\
$sp^3s^*$             & 5 & 4 &  7 & 11 & 11 \\
$sp^3d^2$             & 6 & 3 &  7 & 13 & 13 \\
$sp^3d^5s^*$         & 10 & 7 & 17 & 33 & 33 \\ \hline
$|111\rangle$ ($T_d$) & 4 & 2 & 1 & 1 & 0 \\
$|100\rangle$         & 6 & 2 & 1 & 1 & 0 \\
$|111\rangle$ ($O_h$) & 8 & 3 & 3 & 1 & 0 \\
$|110\rangle$        & 12 & 3 & 1 & 1 & 0 \\
$|110\rangle + |000\rangle$ & 13 & 5 & 1 & 1 & 0 \\
$|111\rangle + |100\rangle$ & 14 & 7 & 5 & 1 & 0 \\
$|111\rangle + |100\rangle + |000\rangle$ & 15 & 11 & 5 & 1 & 0
\end{tabular}
\end{ruledtabular}
\end{table}
Table~\ref{table:freeparam} also lists the properties of other
tight-binding models used in the literature, such as $sp^3s^*$,
\cite{VogHjaDow83} $sp^3d^2$, \cite{ChaAsp90} and
$sp^3d^5s^*$. \cite{Jancu98} The number of parameters listed in this
table is the number permitted by the symmetry of the model, which is
not necessarily the same as that used in any specific implementation
in the literature.

For comparison, the bottom half of Table~\ref{table:freeparam} lists
the properties of several tight-binding models constructed from
symmetrized coordinate eigenkets.  The number of free parameters for
the models generated by $|100\rangle$, $|111\rangle$, and
$|110\rangle$ may be deduced easily from the link geometry results
presented in Appendix \ref{app:link}.  The corresponding numbers for
the compound models (with more than one generator) were determined
using the algorithm in Appendix \ref{app:geometry}.

The most striking feature of Table~\ref{table:freeparam} is the
relative paucity of free parameters in the symmetrized coordinate
approach, which occurs because of the restriction (\ref{eq:Hij0})
imposed by local gauge symmetry.  Several of the symmetrized
coordinate models are direct analogs of conventional models (i.e.,
they have identical symmetry); for example, the $|111\rangle$ ($T_d$)
model corresponds to $sp^3$, and the $|100\rangle$ model corresponds
to $sp^3d^2$ [see Eqs.\ (\ref{eq:Td}) and (\ref{eq:symm_basis})].
However, the number of free parameters in conventional tight-binding
theory grows steadily with distance, whereas in the present theory
there is no coupling beyond second nearest neighbors.

This dearth of adjustable parameters means that the smallest basis
sets do not provide a reliable model for the energy band structure.
For example, in the $|111\rangle$ ($T_d$) model, the splitting of the
bonding and antibonding $s$ states at the $\Gamma$ point is the same
as that of the $p$ states at $\Gamma$ (and that of the $p$ states at
$X$).  In the $|100\rangle$ model, there is no coupling at all between
$p$ orbitals on different atoms at the $\Gamma$ point, so the
splitting of bonding and antibonding states is zero.  Such
difficulties arise primarily because there is only one
nearest-neighbor coupling parameter in these models.

To increase the number of adjustable parameters without an undue
increase in basis size, one must deliberately search for models with
the {\em most complicated} topology available.  A good starting point
is the $|111\rangle$ ($O_h$) basis, which already has 3
nearest-neighbor parameters.  Combining this with a $|100\rangle$
basis raises that number to 4 or 5, depending on the relative values
of the coordinates in the two generators.  This 14-orbital model is a
dramatic improvement over any of the smaller models; however, an extra
$|000\rangle$ site adds substantial extra flexibility without much
change in the basis size.  Thus, the 15-orbital model generated by
$|000\rangle$, $|111\rangle$, and $|100\rangle$ is probably the
smallest basis capable of describing tetrahedral semiconductors
accurately.  In the language of conventional tight-binding theory,
this would be referred to as an $s^3p^6d^5f$ model.

As shown in Table~\ref{table:freeparam}, this model has 17 free
parameters (one of which is just the reference energy).  Specific
definitions for these parameters are given in
Table~\ref{table:modelparam}, which also presents parameter values for
Ge and Si obtained by fitting the band structure of the 15-orbital
model to that given by the nonlocal empirical pseudopotentials of
Chelikowsky and Cohen. \cite{ChCo73L,ChCo74B,ChCo76}
\begin{table*}
\caption{\label{table:modelparam} Independent Hamiltonian parameters
in the 15-orbital model generated by $|0,0,0\rangle$, $|r,r,r\rangle$,
and $|r',0,0\rangle$.}
\begin{ruledtabular}
\begin{tabular}{lldd}
\multicolumn{2}{c}{Parameter} & \multicolumn{2}{c}{Value (Ry)} \\
Symbol & Definition & \multicolumn{1}{c}{Ge} &
 \multicolumn{1}{c}{Si} \\ \hline
$V_a$          & $\langle 0,0,0|H|0,0,0\rangle$ &
 1.22536 & 1.76282 \\
$V_b$          & $\langle r,r,r|H|r,r,r\rangle$ &
 1.18998 & 1.65167 \\
$V_e$          & $\langle -r,r,r|H|-r,r,r\rangle$ &
 1.07211 & 1.17674 \\
$V_f$          & $\langle r',0,0|H|r',0,0\rangle$ &
 1.77902 & 1.98485 \\
$\alpha_{ab}$  & $-\langle 0,0,0|H|r,r,r\rangle$ &
-0.56483 &-0.50749 \\
$\alpha_{ae}$  & $-\langle 0,0,0|H|-r,r,r\rangle$ &
 0.07052 & 0.24801 \\
$\alpha_{af}$  & $-\langle 0,0,0|H|r',0,0\rangle$ &
 0.06400 &-0.07980 \\
$\alpha_{be}$  & $-\langle r,r,r|H|-r,r,r\rangle$ &
 0.03841 & 0.19402 \\
$\alpha_{bf}$  & $-\langle r,r,r|H|r',0,0\rangle$ &
 0.69832 & 0.78464 \\
$\alpha_{ef}$  & $-\langle -r,r,r|H|0,r',0\rangle$ &
 0.31958 & 0.13684 \\
$\alpha_{f\!f}$  & $-\langle r',0,0|H|0,r',0\rangle$ &
-0.43768 &-0.47311 \\
$\beta_{bb}$  & $-\langle r,r,r|H|a-r,a-r,a-r\rangle$ &
 0.74306 & 1.45766 \\
$\beta_{ee}$  & $-\langle -r,r,r|H|a-r,a-r,a+r\rangle$ &
-0.03011 & 0.01537 \\
$\beta_{bf}$  & $-\langle r,r,r|H|a-r',a,a\rangle$ &
-0.07041 &-0.35201 \\
$\beta_{ef}$  & $-\langle -r,r,r|H|a-r',a,a\rangle$ &
 0.28640 & 0.40040 \\
$\beta_{f\!f}$  & $-\langle r',0,0|H|a,a-r',a\rangle$ &
 0.13140 & 0.17401 \\
$\gamma_{ee}$  & $-\langle -r,r,r|H|-r,2a-r,2a-r\rangle$ &
 0.02976 & 0.02025
\end{tabular}
\end{ruledtabular}
\end{table*}
Local Hamiltonian matrix elements are labeled $V$, whereas nonlocal
on-site, nearest-neighbor, and second-nearest-neighbor terms are
denoted $\alpha$, $\beta$, and $\gamma$, respectively.  The subscripts
$a$, $b$, $e$, and $f$ refer to independent sites generated by
$|0,0,0\rangle$, $|r,r,r\rangle$, $|-r,-r,-r\rangle$, and
$|r',0,0\rangle$, respectively.  The labels $a$, $e$, and $f$ are the
Wyckoff labels for sites of different symmetry in the diamond
structure. \cite{Hahn92} The label $b$ is not correct Wyckoff
notation; it actually refers to an independent $e$ site, but the
notation $b$ was used here because these sites lie on the bonds
between atoms.

\begin{figure}
\includegraphics[width=3.375in,keepaspectratio]{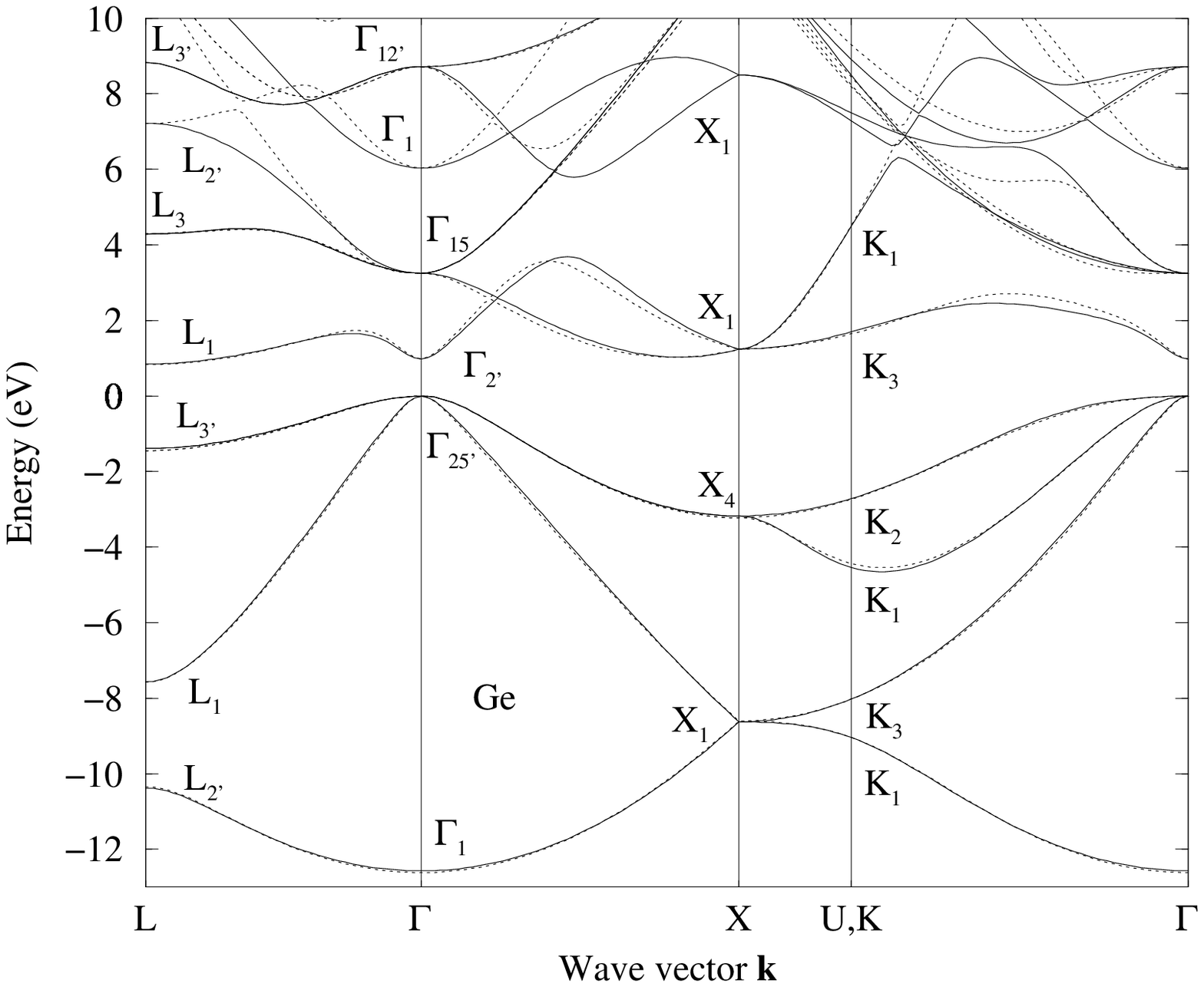}
\includegraphics[width=3.375in,keepaspectratio]{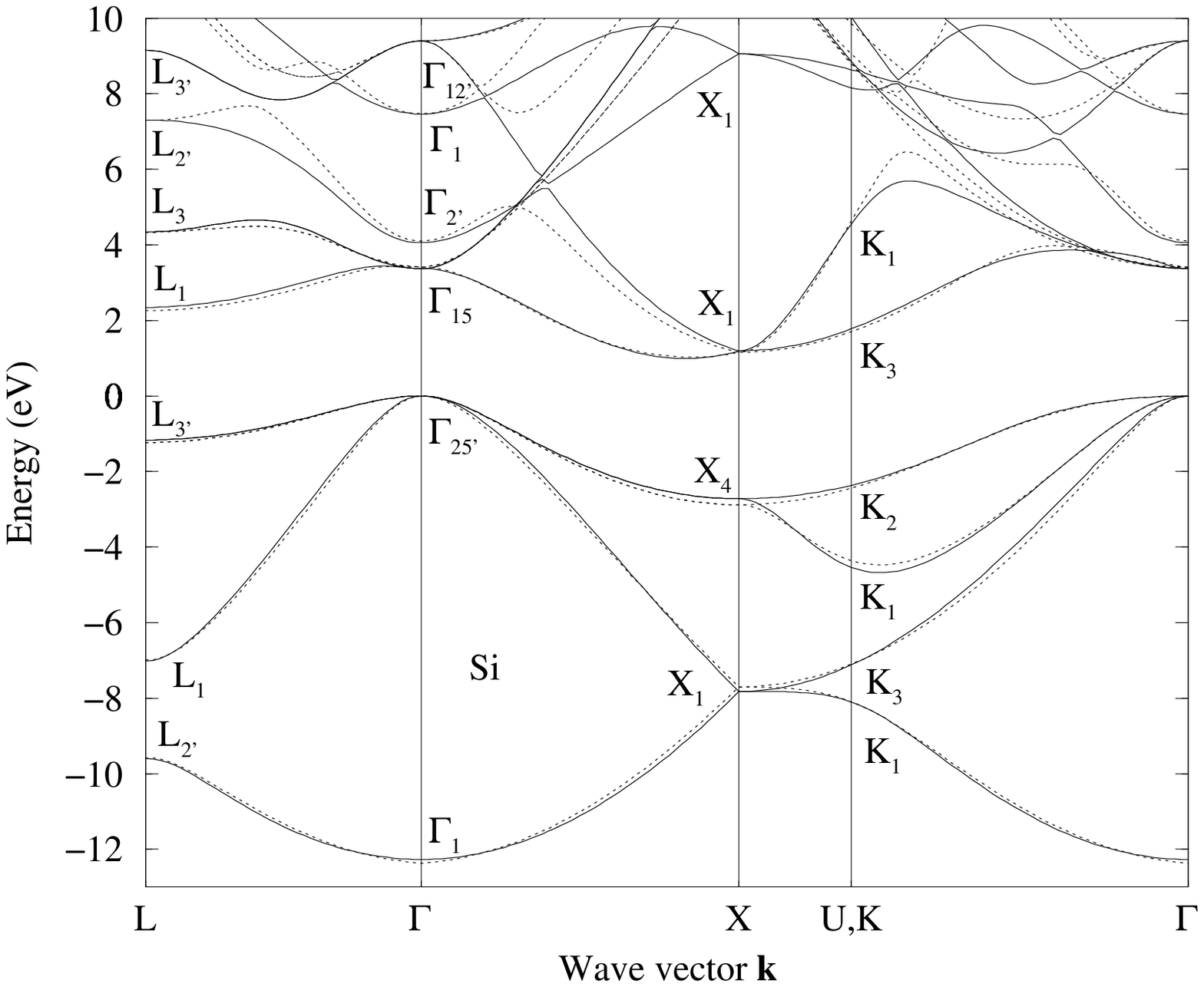}
\caption{\label{fig:GeSi} Energy band structure of germanium and
silicon.  Solid lines: 15-orbital tight-binding model based on
parameters in Table~\ref{table:modelparam}.  Dotted lines: Nonlocal
empirical pseudopotential model of Ref.\ \protect\onlinecite{ChCo76}.
The kinetic-energy cutoff for the latter calculation was $(\mathbf{k}
+ \mathbf{G})^2 \le 21 (2 \pi / a_0)^2$, which corresponds to 113
plane waves at $\Gamma$.}
\end{figure}

The energy band structure calculated from the parameters in
Table~\ref{table:modelparam} is plotted in Fig.\ \ref{fig:GeSi}.
One can see that the 15-orbital model provides a good fit to the
nonlocal pseudopotential bands from the bottom of the valence band to
about 5~eV above the top of the valence band.  Qualitative errors
begin to occur near 9~eV at both the $L$ and $X$ points.  For example,
the $X_1$ level near 9~eV in both figures should occur above 12~eV.
This discrepancy can be eliminated, but the author has not found any
way of doing so without adversely affecting the quality of the overall
fit.

It should be emphasized that the parameters in
Table~\ref{table:modelparam} are presented here merely as ``proof of
concept;'' they are in no way intended as the final word on the
subject, and the author would be surprised if a better set were not
found in the future.  The quantities included in the fitting procedure
were the valence- and conduction-band energy levels at $\Gamma$, $X$,
$L$, and $K$.  Effective masses and deformation potentials were not
considered, and no attempt was made at ensuring transferability.

The main difficulty encountered during the fitting was the lack of any
reliable method for establishing a sound starting point.  Unlike the
case for smaller tight-binding models, the present Hamiltonian has
almost no simple analytical solutions (except for the $\Gamma_{12'}$,
$\Gamma_{12}$, and $X_2$ states, which are relatively unimportant)
that can be used to determine starting values.  The formula
(\ref{eq:Tij}) for the kinetic energy provides a set of
``free-particle'' parameters that is better than nothing, but in a
15-orbital basis, Eq.\ (\ref{eq:Tij}) is a poor approximation to the
continuum kinetic energy (\ref{eq:Tcont}).  After several months and
dozens of different schemes (which still sampled only an infinitesimal
fraction of parameter space), the author was unable to find any method
whose success could honestly be attributed to anything other than
trial and error.  Hence, the development of a robust fitting procedure
remains an unsolved problem.

\subsection{Dielectric function}

As a test of the field-particle coupling in the 15-orbital model, the
imaginary part of the transverse dielectric tensor was calculated from
the formula \cite{GrafVogl95,Bass75}
\begin{eqnarray}
   \epsilon_{2}^{\alpha \beta}(\omega) & = & \frac{4 \pi^2 e^2}{m^2
   \omega^2} \sum_{c,v} \frac{2}{(2 \pi)^3} \int_{\Omega_0^*} \langle
   v \mathbf{k} | p^{\alpha} | c \mathbf{k} \rangle \langle c
   \mathbf{k} | p^{\beta} | v \mathbf{k} \rangle \nonumber \\ && {}
   \times \delta (E_{c\mathbf{k}} - E_{v\mathbf{k}} - \hbar \omega)
   \text{d}^3 \!  k , \label{eq:eps2}
\end{eqnarray}
where $\hbar \omega$ is the photon energy, and $|n \mathbf{k} \rangle$
is an eigenket of $H$ with energy $E_{n \mathbf{k}}$.  The sum covered
the four valence bands $v$ and the seven lowest conduction bands $c$.
The integral was performed using a modified Gilat-Raubenheimer
technique \cite{GilRaub66,SarBr68a,RenHar81} based on 45961
\textbf{k} points in the irreducible part of the Brillouin zone
\cite{Brust64,Corn84} (representing 2048000
points in the full Brillouin zone).  The energy interval for this
calculation was 1~meV.

To reveal more clearly the physical meaning of the calculated spectra,
the same method was used to calculate the joint density of states
function
\begin{equation}
   J(E) = \sum_{c,v} \frac{2 \Omega_0}{(2 \pi)^3} \int_{\Omega_0^*}
   \delta (E_{c\mathbf{k}} - E_{v\mathbf{k}} - E) \text{d}^3 \!  k .
   \label{eq:jdos}
\end{equation}
The dielectric function differs from $J(E)$ in that each transition is
weighted by the oscillator strength
\begin{equation}
   f_{cv}^{\alpha\beta} = \frac{2 \langle v \mathbf{k} | p^{\alpha} |
   c \mathbf{k} \rangle \langle c \mathbf{k} | p^{\beta} | v
   \mathbf{k} \rangle}{m (E_{c\mathbf{k}} - E_{v\mathbf{k}})}
   . \label{eq:fcv}
\end{equation}
One may therefore use $\epsilon_2$ and $J$ to define an average
oscillator strength at each energy by
\begin{equation}
   F^{\alpha\beta} (\hbar \omega) = \frac{m \omega \Omega_0}{2 \hbar
   \pi^2 e^2} \frac{\epsilon_2^{\alpha\beta} (\omega)}{J(\hbar
   \omega)} . \label{eq:Fave}
\end{equation}
In cubic crystals, the tensors (\ref{eq:eps2}) and (\ref{eq:Fave})
reduce to scalars: $\epsilon_2^{\alpha\beta}(\omega) =
\epsilon_2(\omega) \delta_{\alpha\beta}$.

\begin{figure}
\includegraphics[width=3.375in,keepaspectratio]{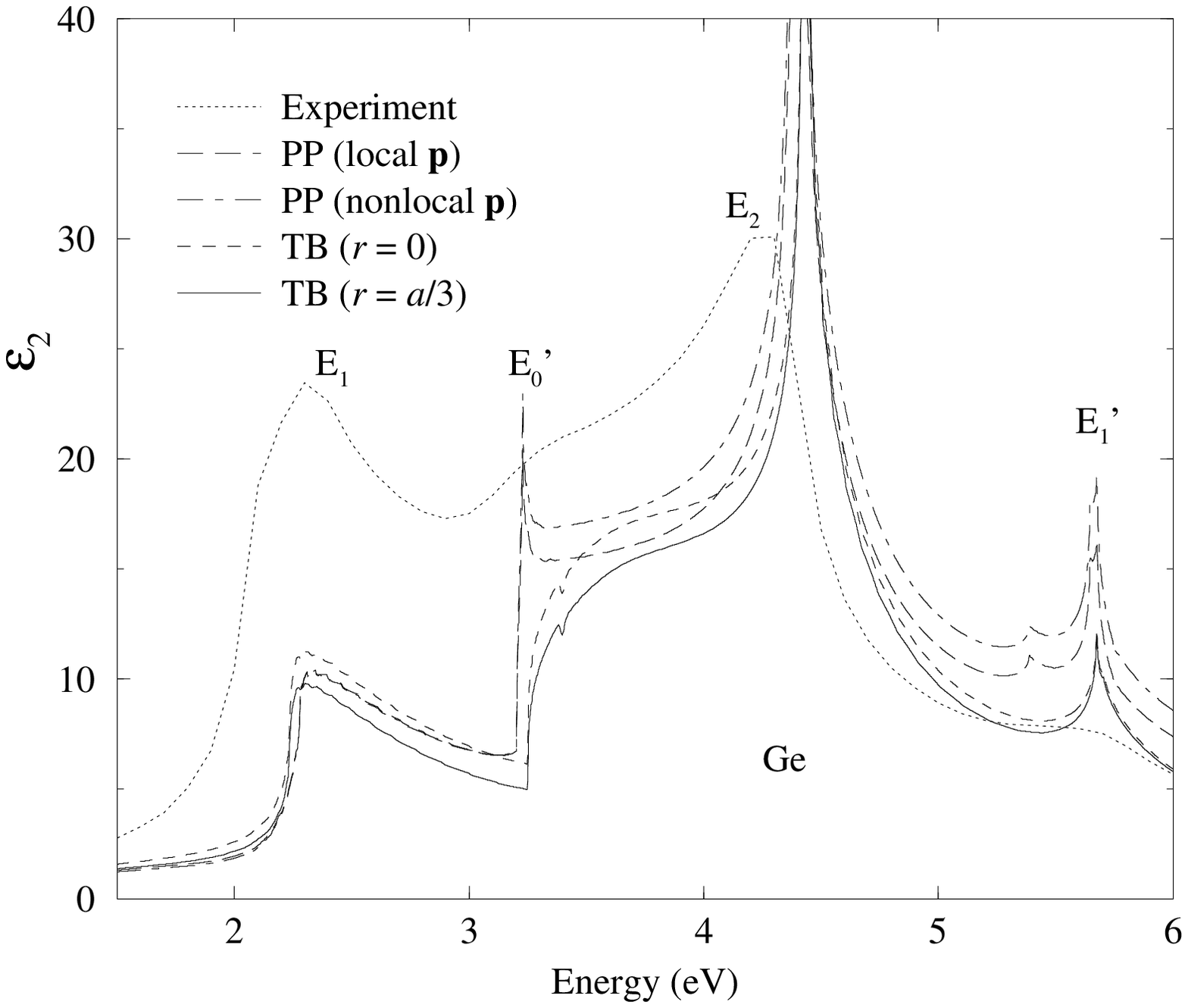}
\includegraphics[width=3.375in,keepaspectratio]{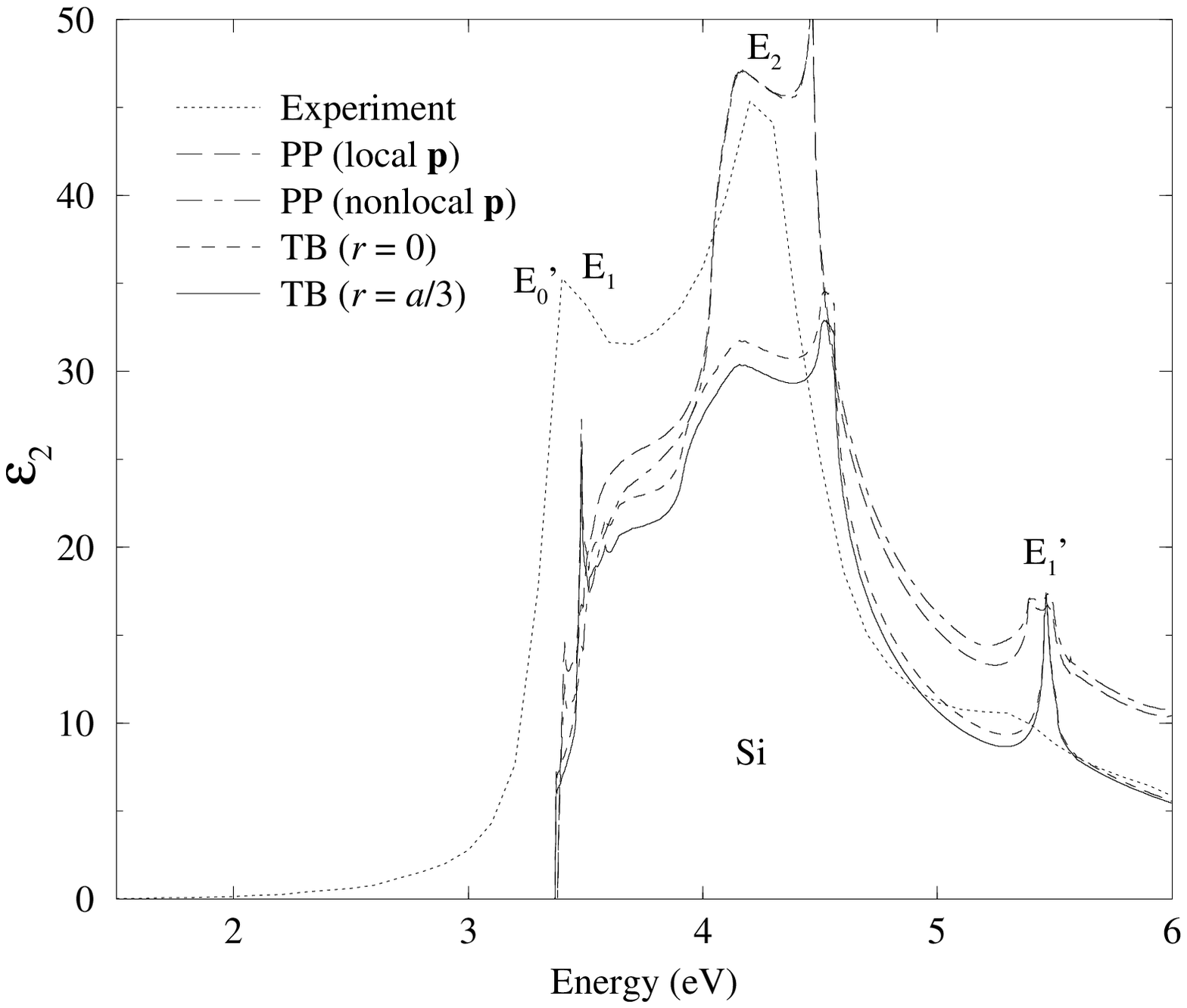}
\caption{\label{fig:eps2} Imaginary part of the transverse dielectric
function of germanium and silicon.  Dotted line: Experimental data
from Ref.\ \protect\onlinecite{AspStu83}.  Long dashed line: Nonlocal
pseudopotential model of Ref.\ \protect\onlinecite{ChCo76} with
canonical (local) momentum $\mathbf{p} = -i \hbar \nabla$.  Dot-dashed
line: Nonlocal pseudopotential model of Ref.\
\protect\onlinecite{ChCo76} with kinematic (nonlocal) momentum
$\mathbf{p} = (m / i \hbar) [\mathbf{x}, H]$.  Short dashed line:
15-orbital tight-binding model from Table~\ref{table:modelparam} with
$r \rightarrow 0$ and $r' \rightarrow 0$.  Solid line: 15-orbital
tight-binding model from Table~\ref{table:modelparam} with $r =
\frac13 a$ and $r' = \sqrt{2} r$.}
\end{figure}

The calculated dielectric function $\epsilon_2(\omega)$ for Ge
and Si is plotted in Fig.\ \ref{fig:eps2}.
This figure compares experimental data \cite{AspStu83} for the
dielectric function with the values given by Eq.\ (\ref{eq:eps2}) for
(i)~the nonlocal pseudopotential model of Chelikowsky and Cohen,
\cite{ChCo73L,ChCo74B,ChCo76} and (ii)~the 15-orbital tight-binding
model of Table~\ref{table:modelparam}.  For each model, two plots of
$\epsilon_2(\omega)$ are given, corresponding to two different
expressions for the momentum operator $\mathbf{p}$.

In pseudopotential calculations, optical properties are usually
calculated from $\mathbf{A} \cdot \mathbf{p}$ coupling with
$\mathbf{p} = -i \hbar \nabla$. \cite{ChCo76} However, if the
pseudopotential is nonlocal, this coupling is not gauge invariant; the
correct linear coupling is given instead by the kinematic momentum
$\mathbf{p} = (m / i \hbar) [ \mathbf{x}, H ]$.  \cite{IsChLouie01}
Since pseudopotential calculations are usually performed in a
plane-wave basis, a more convenient expression for the kinematic
momentum is given by Eq.\ (\ref{eq:pk}) (which is valid for both
discrete and continuous coordinates $\mathbf{x}$).

In the present tight-binding theory, the kinematic momentum is given
by Eq.\ (\ref{eq:pH_tb}).  The two tight-binding functions plotted in
Fig.\ \ref{fig:eps2} correspond to two different choices of the
intra-atomic coordinates $\mathbf{x}_{\alpha}$, which are determined
by the parameters $r$ and $r'$ in Table~\ref{table:modelparam}.  One
choice was the limit $r \rightarrow 0$, $r' \rightarrow 0$, which is
equivalent to the zero-parameter model of Eqs.\ (\ref{eq:x0}) and
(\ref{eq:Peierls}).  The other, more physically realistic choice was
$r = \frac13 a$ and $r' = \sqrt{2} r$.  The value $r = \frac13 a$ was
chosen because it yields equidistant lattice sites along the bond
directions $\langle 111 \rangle$.  The value $r' = \sqrt{2} r$ was
used because a somewhat larger value (e.g., $1.5r$) breaks the link
$\alpha_{f\!f}$ in Table \ref{table:modelparam}, whereas a somewhat
smaller value (e.g., $1.3r$) breaks the links $\beta_{bf}$ and
$\beta_{f\!f}$ (while simultaneously forming a new link $\beta_{be}$).
These values of $r$ and $r'$ also generated successful starting values
for some of the parameters in Table \ref{table:modelparam} (although
the final fitted parameters were not very close to the starting
values).

Several conclusions may be drawn from Fig.\ \ref{fig:eps2}.  The first
is that, within a given model (pseudopotential or tight binding), the
choice of momentum operator does not have much numerical significance
for the present calculation.  This was to be expected on physical
grounds, since the intra-atomic coordinate $\mathbf{x}_{\alpha}$ (in
the tight-binding model) and the nonlocal part of the momentum (in the
pseudopotential model) both lead to polarization effects {\em within}
the atom.  Yet it is well known that the bonds between atoms are much
easier to polarize than the atoms themselves. \cite{Har89,Har99}
Hence, in a bulk semiconductor, intra-atomic effects yield only a
minor numerical correction.  This conclusion should remain valid in
any system where the states are extended, but it may break down in
systems where localized states are important. \cite{Cruz99,PedKri01}

The nonlocal part of the pseudopotential momentum tends to increase
$\epsilon_2(\omega)$ in most frequency ranges, but it sometimes has
the opposite effect (see, e.g., the region below 4~eV in Si).
However, the intra-atomic coupling in the tight-binding model always
decreases $\epsilon_2(\omega)$.  This may be understood by noting that
the dominant nonlocal term in the Hamiltonian of Table
\ref{table:modelparam} is the coupling $\beta_{bb}$ along the bond
between nearest neighbors.  Increasing the value of $r$ decreases the
distance between $b$ sites on neighboring atoms, thereby decreasing
the momentum matrix in Eq.\ (\ref{eq:pH_tb}).  This tends to increase
(slightly) the discrepancy between the tight-binding and
pseudopotential dielectric functions.  It is possible, however, that a
different parametrization of the Hamiltonian might yield different
results.

\begin{figure}
\includegraphics[width=3.375in,keepaspectratio]{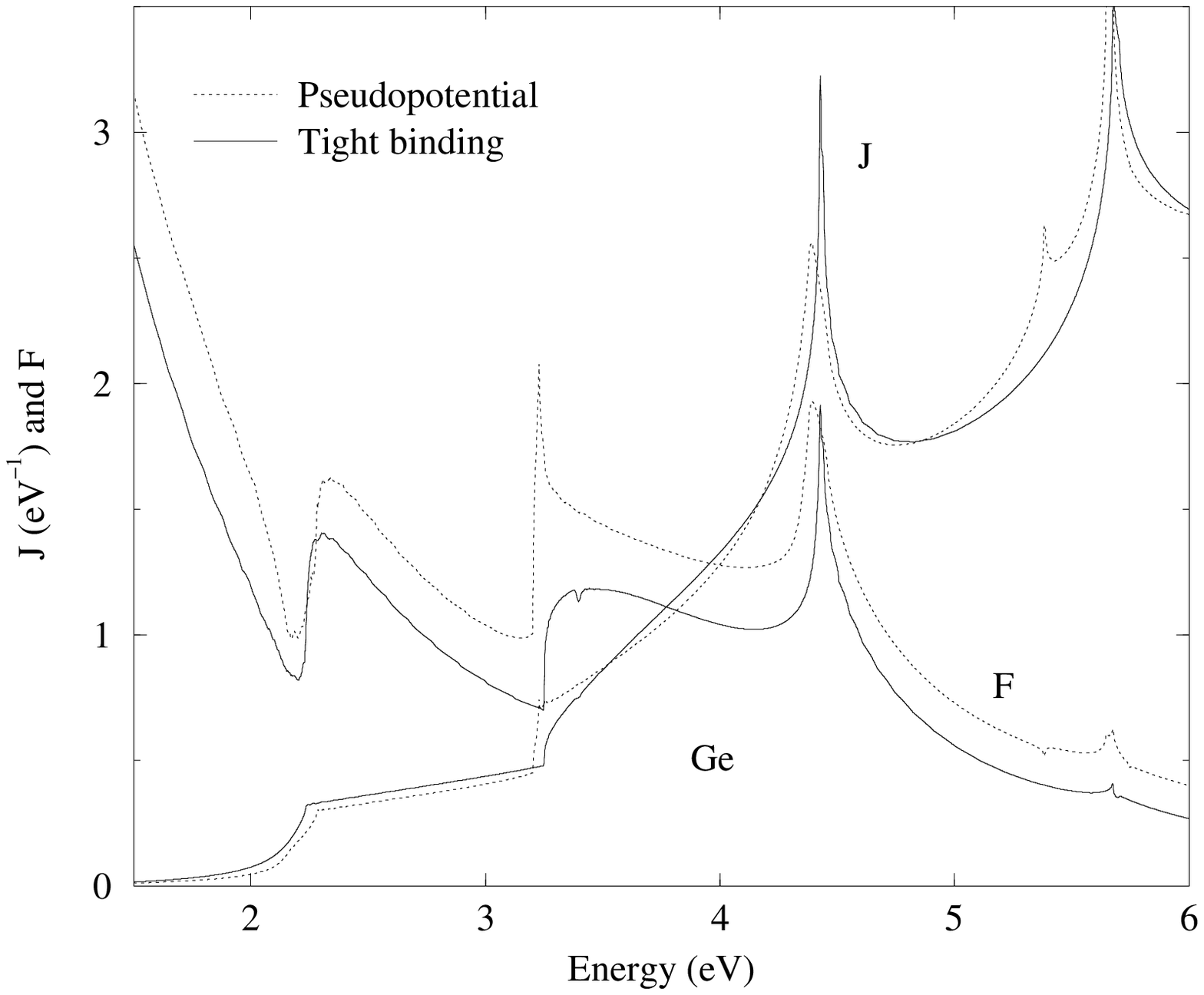}
\includegraphics[width=3.375in,keepaspectratio]{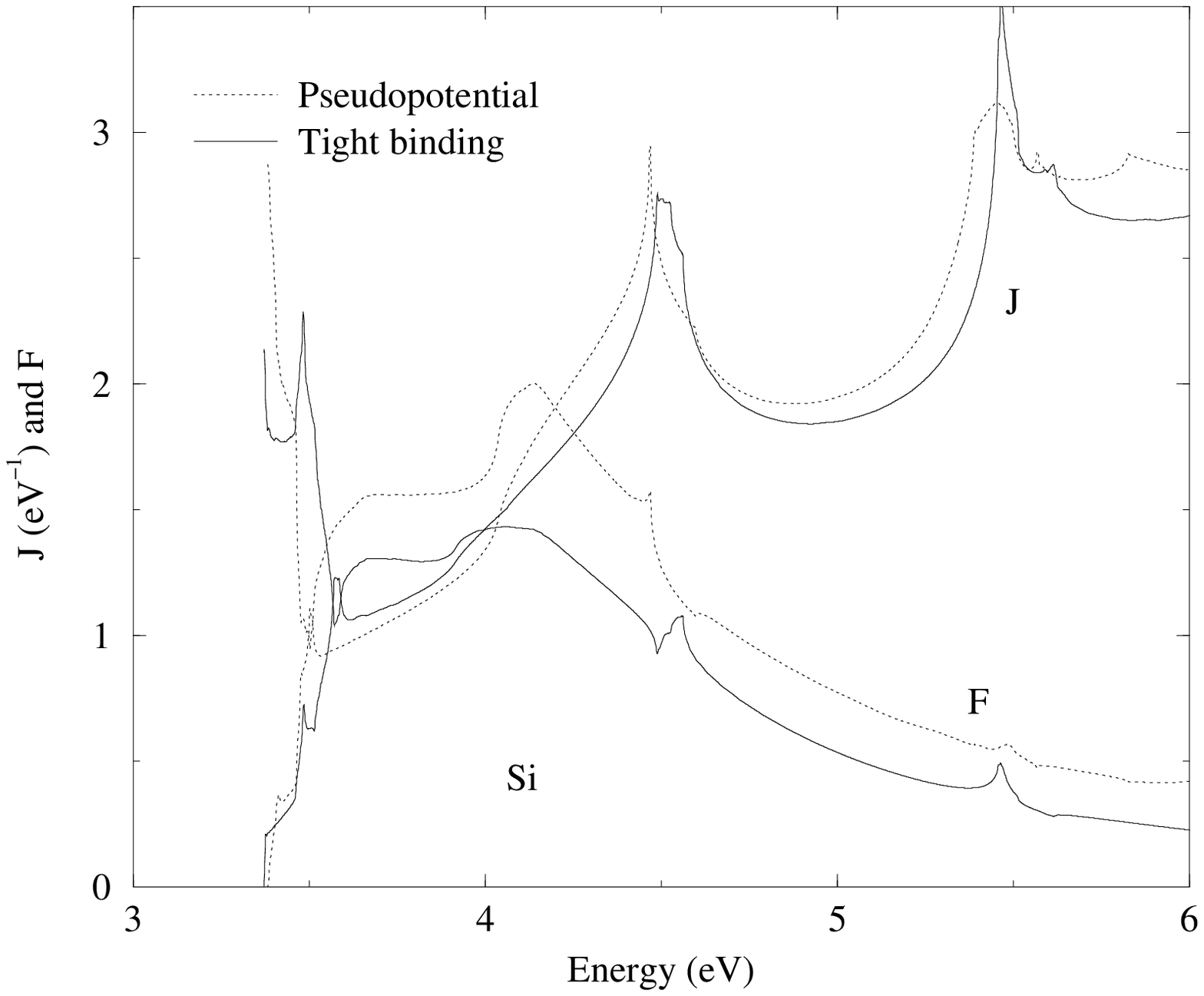}
\caption{\label{fig:jdos} Joint density of states $J$ and average
oscillator strength $F$ for Ge and Si.  Dotted line: Nonlocal
pseudopotential model of Ref.\ \protect\onlinecite{ChCo76} with
kinematic momentum $\mathbf{p} = (m / i \hbar) [\mathbf{x}, H]$.
Solid line: 15-orbital tight-binding model from
Table~\ref{table:modelparam} with $r = \frac13 a$ and $r' = \sqrt{2}
r$.}
\end{figure}

The tight-binding and pseudopotential dielectric functions are quite
similar in Ge, but there is a significant discrepancy at the $E_2$
peak in Si.  The reason for the difference between the models is
apparent from the joint density of states $J$ and average oscillator
strength $F$ plotted in Fig.\ \ref{fig:jdos}.
This figure shows that the tight-binding $J$ is very accurate in Ge
and somewhat less so in Si, as might have been expected from the
quality of the fitted energy bands in Fig.\ \ref{fig:GeSi}.  However,
the tight-binding model underestimates the oscillator strength over
almost the entire frequency range shown, typically by about 20\%.  The
difference is most pronounced between 4 and 4.5~eV in Si, where it
exceeds 30\%.  When combined with a slight underestimate of $J$ in the
same region, this leads to the discrepancy in the $E_2$ peak noted
above.

The $E_2$ peak in Si is associated with a volume in $\mathbf{k}$ space
near $(0.9,0.1,0.1)2\pi/a_0$, \cite{ChCo76} which is close to the $X$
point.  Thus, the physical reason for the error in the $E_2$ peak is
probably the spurious $X_1$ conduction band near 9~eV in Fig.\
\ref{fig:GeSi}.  Because this band is too low in energy, it mixes more
strongly with the lowest $X_1$ conduction band in the tight-binding
model than it does in the pseudopotential model.  This change in the
wave function causes a corresponding change in oscillator strength.
Thus, the majority of the error in the Si $E_2$ peak would likely be
eliminated if one could find an improved parameter set that raises the
energy of the upper $X_1$ conduction band.

It is less clear whether the systematic underestimate of oscillator
strength at all frequencies could be resolved by changing the
Hamiltonian parameters.  Oscillator strength was not included in the
present fitting routine, so it is possible that with specific
attention to this feature, one could improve the oscillator strength
while maintaining the quality of the joint density of states.
However, it is also possible that such an underestimate is a
fundamental limitation imposed by the small basis size in the
tight-binding model.  Thus, at present, the 15-orbital model is
capable of providing semiquantitative predictions of oscillator
strength that reproduce all of the major trends exhibited by the
pseudopotential model.  Whether future developments bring it into
precise quantitative agreement remains to be seen.

Finally, it is worth noting that in Fig.\ \ref{fig:eps2}, the
calculated $E_1$ peak (between 2 and 3~eV) for Ge is considerably less
than the experimental value for both the pseudopotential and
tight-binding models.  Chelikowsky and Cohen have attributed this
discrepancy to the neglect of exciton effects. \cite{ChCo73L} However,
a recent first-principles calculation of $\epsilon_2(\omega)$ that
includes electron-hole interactions \cite{BeShBo98} shows that the
contribution from excitons in Ge is not large enough to fill the gap
in Fig.\ \ref{fig:eps2}.  Thus, the empirical pseudopotential for Ge
\cite{ChCo73L,ChCo76} probably needs further adjustment to increase
the $E_1$ peak.

\section{Discussion and conclusions}

\label{sec:conclusion}

This paper has shown that intra-atomic optical transitions can be
incorporated into tight-binding theory in a gauge-invariant way if the
coordinate representation is taken as fundamental.  Orthogonal
atomic-like orbitals can be constructed from symmetrized coordinate
eigenkets, and the coupling to electromagnetic fields can then be
described using lattice gauge theory.  A model based on 15 such
orbitals per atom is capable of describing the most important features
of the tetrahedral semiconductors Ge and Si.  This basis is slightly
larger than existing 10-orbital models, \cite{Jancu98} but it has the
advantages of (i)~gauge invariance and (ii)~providing an explicit wave
function for the electron.  A larger basis is needed in the present
theory because the restrictions imposed by local gauge symmetry reduce
the number of available Hamiltonian fitting parameters.

The field-particle coupling derived here is similar to that given by
the Peierls substitution,
\cite{Pei33,GrafVogl95,Dum98,Boyk01,IsChLouie01,BoykVogl02} but the
field-induced phase factor appears in the coordinate representation
rather than the tight-binding representation.  Thus, the present
formalism includes intra-atomic coupling not present in the Peierls
substitution.  \citet*{IsChLouie01} have recently presented a
derivation of the Peierls phase for nonlocal Hamiltonians in a
continuous coordinate representation.  They have argued that this
derivation justifies the use of the Peierls substitution in
tight-binding theory.  However, their derivation cannot be
extrapolated to tight-binding theory, because ordinary $\mathbf{A}
\cdot \mathbf{p}$ coupling gives rise to intra-atomic interactions
that are not included in the Peierls substitution.  The existence of
such interactions was not considered in the tight-binding theory of
Ref.\ \onlinecite{IsChLouie01}.

It is interesting to consider whether there are any other ways of
incorporating local gauge symmetry into tight-binding theory.  One
possibility is to work directly in the usual tight-binding
representation [see, e.g., Eq.\ (\ref{eq:Peierls})], where the basis
kets are labeled by the symmetry of the orbital and the position of
the atom.  If gauge symmetry is to be applied in this basis, the
coordinate operator must be diagonal, hence all intra-atomic matrix
elements must be set to zero [as in Eq.\ (\ref{eq:x0})].  One could
then introduce an Abelian $U(1)$ gauge field on this lattice using the
approach described above in Sec.\ \ref{sec:gauge}.  The results would
be identical to those found in Sec.\ \ref{sec:gauge}, except that the
phase factor in the Hamiltonian (\ref{eq:Hfield}) would be applied in
the tight-binding representation rather than the coordinate
representation.  Hence, this approach would constitute a
``derivation'' of the Peierls phase (\ref{eq:Peierls}).  Such a
derivation would eliminate the ambiguity associated with the choice of
path \cite{GrafVogl95} in Eq.\ (\ref{eq:Peierls}). (Other techniques
for eliminating path ambiguity are described in Refs.\
\onlinecite{IsChLouie01} and \onlinecite{Boyk01}.)

The problem with this approach lies in its treatment of the coordinate
operator.  In the theory described in Sec.\ \ref{sec:gauge}, when the
basis size is increased, the eigenvalue spectrum of the coordinate
operator remains nondegenerate, tending (in the limit of infinite
basis dimensions) toward a continuous spectrum.  However, if the
coordinate operator is required to be diagonal in the tight-binding
basis, its eigenvalue spectrum is always degenerate, tending (in the
limit of infinite basis dimensions) toward a discrete spectrum with
infinite degeneracy.  Hence, any tight-binding theory that is either
based on or equivalent to the Peierls substitution cannot reproduce
the correct continuum limit of the coordinate operator.

As a generalization of the above approach, one might also consider
introducing a non-Abelian gauge field
\cite{YangMills54,WuYang75,Mor83,Guidry91,Wil74,KogSus75,Rebbi83,%
Rothe97,ChrFriLee82a,ChrFriLee82b,ChrFriLee82c} in the tight-binding
basis.  The idea would be to treat the tight-binding electron as a new
type of ``elementary particle'' with some internal degrees of freedom
(corresponding to the symmetry labels of the atomic orbitals) that are
coupled to the gauge field.  In this way, one might hope to reproduce
the effects of intra-atomic coupling while remaining in the
tight-binding basis.  There are, however, numerous difficulties with
this approach.

First, the lattice sites in lattice gauge theory represent states of
the same particle at different positions.  Hence, these states are
identical apart from their positions.  However, in tight-binding
theory the atoms are generally not the same.  Second, the field
equations for a non-Abelian gauge field are intrinsically nonlinear,
because the field carries its own charge and is coupled directly to
itself (i.e., it is self-radiating).  It is therefore difficult to
imagine how such a field could reproduce ordinary electromagnetism, in
which the field has no charge, and nonlinearities arise only from
interactions with matter.  Third, one would need to define a new gauge
field theory every time one added new orbitals to the model, and every
one of these non-Abelian field theories would need to reproduce the
results of Abelian electromagnetic theory.  Finally, the coordinate
operator in this approach would still have a discrete, degenerate
eigenvalue spectrum.

Thus, it appears that the present approach---that is, an Abelian
$U(1)$ gauge field in the coordinate representation---is the only
gauge-invariant method for including electromagnetic fields in
empirical tight-binding theory that tends toward the correct continuum
limit as the basis dimensions are increased.  In this case, the only
way that the essential structure of the theory can be modified is to
change the topology of the system so as to increase the number of
links between lattice sites.  This would increase the number of free
parameters in the Hamiltonian, thereby permitting a reduction in basis
size.  Such a modification would clearly be beneficial, but it is not
obvious that there exists any alternative topology for general
lattices that is capable of reproducing continuum electromagnetism
unambiguously.  Hence, this possibility will not be explored further
here.

\begin{acknowledgments}
I am grateful to Peter Vogl, Tim Boykin, and Tai Kai Ng for helpful
discussions.  This work was supported by Hong Kong RGC Grant No.\
HKUST6139/00P.
\end{acknowledgments}

\appendix

\section{Symmetrized orbitals}

\label{app:basis}

This section presents symmetrized orbitals obtained by applying the
symmetry operations of the cubic group $O_h$ to the coordinate
eigenkets $|100\rangle$ and $|110\rangle$.  [The orbitals for
$|111\rangle$ may be found in Eq.\ (\ref{eq:ccc_tb}).] For
$|100\rangle$, if the basis kets are ordered as $\{ | 100 \rangle$, $|
010 \rangle$, $| 001 \rangle$, $| \bar{1}00 \rangle$, $| 0\bar{1}0
\rangle$, $| 00\bar{1}\rangle \}$, then the symmetrized orbitals are
\begin{eqnarray}
   | \Gamma_1 \rangle & = & | s \rangle \nonumber \\ & = &
   \textstyle\frac{1}{\sqrt{6}} (1,1,1,1,1,1) ,
   \nonumber \\
   | \Gamma_{15}^{z} \rangle & = & | p_z \rangle  \nonumber \\ & = &
   \textstyle\frac{1}{\sqrt{2}} (0,0,1,0,0,-1) ,
   \nonumber \\
   | {\Gamma}_{12}^{a} \rangle & = & | d_{2z^2-x^2-y^2} \rangle 
    \nonumber \\ & = &
   \textstyle\frac{1}{2\sqrt{3}} (-1,-1,2,-1,-1,2) ,
   \nonumber \\
   | {\Gamma}_{12}^{b} \rangle & = & | d_{x^2-y^2} \rangle 
   \nonumber \\ & = &
   \textstyle\frac{1}{2} (1,-1,0,1,-1,0) .
\end{eqnarray}
For $|110\rangle$, if the basis kets are ordered as $\{ | 011
\rangle$, $| 01\bar{1} \rangle$, $| 0\bar{1}1 \rangle$, $|
0\bar{1}\bar{1} \rangle$, $| 101 \rangle$, $| 10\bar{1} \rangle$, $|
\bar{1}01 \rangle$, $| \bar{1}0\bar{1} \rangle$, $| 110 \rangle$, $|
1\bar{1}0 \rangle$, $| \bar{1}10 \rangle$, $| \bar{1}\bar{1}0 \rangle
\}$, then the symmetrized orbitals are
\begin{eqnarray}
   | \Gamma_1 \rangle & = & | s \rangle \nonumber \\ & = &
   \textstyle\frac{1}{2\sqrt{3}} (1,1,1,1,1,1,1,1,1,1,1,1) ,
   \nonumber \\
   | \Gamma_{15}^{z} \rangle & = & | p_z \rangle  \nonumber \\ & = &
   \textstyle\frac{1}{2\sqrt{2}} (1,-1,1,-1,1,-1,1,-1,0,0,0,0) ,
   \nonumber \\
   | {\Gamma}_{12}^{a} \rangle & = & | d_{2z^2-x^2-y^2} \rangle 
    \nonumber \\ & = &
   \textstyle\frac{1}{2\sqrt{6}} 
   (-1,-1,-1,-1,-1,-1,-1,-1,2,2,2,2) ,
   \nonumber \\
   | {\Gamma}_{12}^{b} \rangle & = & | d_{x^2-y^2} \rangle 
   \nonumber \\ & = &
   \textstyle\frac{1}{2\sqrt{2}} 
   (1,1,1,1,-1,-1,-1,-1,0,0,0,0) ,
    \nonumber \\
   | \Gamma_{25'}^{xy} \rangle & = & | d_{xy} \rangle
   \nonumber \\ & = &
   \textstyle\frac{1}{2} (0,0,0,0,0,0,0,0,1,-1,-1,1) ,
   \nonumber \\
   | \Gamma_{25}^{c} \rangle & = & | f_{z(x^2-y^2)} \rangle 
   \nonumber \\ & = &
   \textstyle\frac{1}{2\sqrt{2}} (-1,1,-1,1,1,-1,1,-1,0,0,0,0) .
\end{eqnarray}
For the triply degenerate representations $\Gamma_{15}$,
$\Gamma_{25'}$, and $\Gamma_{25}$, only one representative orbital is
given; the others may be obtained from cyclic permutations of $x$,
$y$, and $z$.

\section{Geometry of Voronoi polyhedra}

\label{app:geometry}

This appendix presents an algorithm for calculating the geometry of
the Voronoi polyhedra associated with a given set of nodes
$\mathbf{x}_i$.  The basic element in this algorithm is a procedure
for finding the edges of the polyhedra.  An edge of a Voronoi
polyhedron is a finite line segment consisting of points that are
closer to three (or more) nodes than to any other nodes.  The first
step is therefore to determine the equation defining this line.

Any three noncollinear points $\mathbf{x}_i$, $\mathbf{x}_j$, and
$\mathbf{x}_k$ define a plane whose normal is the vector
\begin{eqnarray}
   \mathbf{n}_{ijk} & = & \mathbf{d}_{ji} \times \mathbf{d}_{ki}
   \nonumber \\ & = & \mathbf{x}_i \times \mathbf{x}_j + \mathbf{x}_j
   \times \mathbf{x}_k + \mathbf{x}_k \times \mathbf{x}_i ,
   \label{eq:nijk}
\end{eqnarray}
where $\mathbf{d}_{ji} = \mathbf{x}_{j} - \mathbf{x}_{i}$.  This plane
is the set of points $\mathbf{x}$ satisfying
\begin{equation}
   \mathbf{n}_{ijk} \cdot (\mathbf{x} - \mathbf{x}_i) = 0
   . \label{eq:plane}
\end{equation}
The line consisting of all points $\mathbf{x}$ equidistant from
$\mathbf{x}_i$, $\mathbf{x}_j$, and $\mathbf{x}_k$ may therefore be
written as
\begin{equation}
   \mathbf{x} = \mathbf{x}_{ijk} + \lambda \hat{\mathbf{n}}_{ijk} ,
   \label{eq:line_ijk}
\end{equation}
where $\lambda$ is a real parameter, $\hat{\mathbf{n}}_{ijk} =
\mathbf{n}_{ijk} / n_{ijk}$, and $\mathbf{x}_{ijk}$ is the point in
the plane (\ref{eq:plane}) equidistant from $\mathbf{x}_i$,
$\mathbf{x}_j$, and $\mathbf{x}_k$.  To determine this point, note
that points $\mathbf{x}$ equidistant from $\mathbf{x}_i$ and
$\mathbf{x}_j$ satisfy
\begin{equation}
   (\mathbf{x}_{j} - \mathbf{x}_{i}) \cdot [ \mathbf{x} -
   \textstyle\frac12 (\mathbf{x}_{j} + \mathbf{x}_{i}) ] = 0
   . \label{eq:plane_ij}
\end{equation}
The point $\mathbf{x}_{ijk}$ therefore satisfies the three equations
\begin{equation}
   \mathbf{a}_{r} \cdot \mathbf{x}_{ijk} = c_{r} \qquad (r = 1,2,3)
   \label{eq:xijk0}
\end{equation}
in which
\begin{equation}
   \mathbf{a}_1 = \mathbf{d}_{ji} , \qquad \mathbf{a}_2 =
   \mathbf{d}_{ki} , \qquad \mathbf{a}_3 = \mathbf{n}_{ijk} ,
\end{equation}
which may be viewed as a set of oblique (more specifically,
monoclinic) basis vectors, and
\begin{eqnarray}
   c_1 & = & \textstyle\frac12 (x_j^2 - x_i^2) , \nonumber \\
   c_2 & = & \textstyle\frac12 (x_k^2 - x_i^2) , \nonumber \\
   c_3 & = & \mathbf{n}_{ijk} \cdot \mathbf{x}_i \nonumber \\
       & = & \mathbf{x}_i \cdot (\mathbf{x}_j \times \mathbf{x}_k) .
\end{eqnarray}
The solution to Eqs.\ (\ref{eq:xijk0}) is given by
\begin{equation}
   \mathbf{x}_{ijk} = \sum_{s=1}^{3} c_s \mathbf{b}_s ,
   \label{eq:xijk1}
\end{equation}
where $\mathbf{b}_s$ is a reciprocal basis vector satisfying
$\mathbf{a}_r \cdot \mathbf{b}_s = \delta_{rs}$; e.g.,
\begin{equation}
   \mathbf{b}_1 = \frac{\mathbf{a}_2 \times \mathbf{a}_3}{\mathbf{a}_1
   \cdot (\mathbf{a}_2 \times \mathbf{a}_3)}.
\end{equation}
Now since $\mathbf{a}_1 \cdot (\mathbf{a}_2 \times \mathbf{a}_3) =
n_{ijk}^2$, the vectors $\mathbf{b}_s$ are given explicitly by
\begin{eqnarray}
   \mathbf{b}_1 & = & [\mathbf{d}_{ji} (d_{ki}^2) - \mathbf{d}_{ki}
   (\mathbf{d}_{ji} \cdot \mathbf{d}_{ki})] / n_{ijk}^2 , \nonumber \\
   \mathbf{b}_2 & = & [\mathbf{d}_{ki} (d_{ji}^2) - \mathbf{d}_{ji}
   (\mathbf{d}_{ji} \cdot \mathbf{d}_{ki})] / n_{ijk}^2 , \\
   \mathbf{b}_3 & = & \mathbf{n}_{ijk} / n_{ijk}^2 . \nonumber
\end{eqnarray}
Equation (\ref{eq:xijk1}) may then be rearranged in the more symmetric
form
\begin{widetext}
\begin{equation}
   \mathbf{x}_{ijk} = \frac{\mathbf{x}_{i} [d_{jk}^2 (\mathbf{d}_{ji}
   \cdot \mathbf{d}_{ki})] + \mathbf{x}_{j} [d_{ki}^2 (\mathbf{d}_{kj}
   \cdot \mathbf{d}_{ij})] + \mathbf{x}_{k} [d_{ij}^2 (\mathbf{d}_{ik}
   \cdot \mathbf{d}_{jk})]}{2 n_{ijk}^2} . \label{eq:xijk2}
\end{equation}
\end{widetext}
This result, together with Eq.\ (\ref{eq:line_ijk}), defines the line
equidistant from nodes $\mathbf{x}_i$, $\mathbf{x}_j$, and
$\mathbf{x}_k$.

The next step is to determine whether any segment of this line forms
an edge of a Voronoi polyhedron.  Points on such a segment must lie
closer to $\mathbf{x}_i$, $\mathbf{x}_j$, and $\mathbf{x}_k$ than to
any other node $\mathbf{x}_l$.  For each node $\mathbf{x}_l$, one
calculates
\begin{equation}
   \alpha_l = \mathbf{d}_{il} \cdot \hat{\mathbf{n}}_{ijk} .
\end{equation}
If $\alpha_l = 0$, then $\mathbf{x}_l$ lies in the plane
(\ref{eq:plane}).  In this case, if $|\mathbf{x}_{l} -
\mathbf{x}_{ijk}| < |\mathbf{x}_{i} - \mathbf{x}_{ijk}|$, then no
portion of the line (\ref{eq:line_ijk}) forms an edge of a Voronoi
polyhedron.  On the other hand, if $|\mathbf{x}_{l} -
\mathbf{x}_{ijk}| \ge |\mathbf{x}_{i} - \mathbf{x}_{ijk}|$, then the
line (\ref{eq:line_ijk}) may form an edge (depending on the position
of the other nodes $\mathbf{x}_{l'}$).

If $\alpha_l \neq 0$, then $\mathbf{x}_l$ does not lie in the plane
(\ref{eq:plane}).  In this case, points on the line
(\ref{eq:line_ijk}) that are closer to $\mathbf{x}_i$ than to
$\mathbf{x}_l$ satisfy [cf.\ Eq.\ (\ref{eq:plane_ij})]
\begin{equation}
   \mathbf{d}_{il} \cdot ( \mathbf{x}_{ijk} + \lambda
   \hat{\mathbf{n}}_{ijk} - \mathbf{x}_{il} ) > 0 ,
\end{equation}
in which $\mathbf{x}_{il} = \frac12 (\mathbf{x}_{i} +
\mathbf{x}_{l})$.  Hence, the position of the point equidistant from
$\mathbf{x}_i$, $\mathbf{x}_j$, $\mathbf{x}_k$, and $\mathbf{x}_l$ is
given by the following value of the parameter $\lambda$ in Eq.\
(\ref{eq:line_ijk}):
\begin{equation}
   \lambda_l = \frac{\mathbf{d}_{il} \cdot (\mathbf{x}_{il} -
   \mathbf{x}_{ijk})}{\alpha_l} .
\end{equation}
One may then define
\begin{equation}
   \lambda_{\text{min}} = \text{max}(\lambda_l | \alpha_l > 0)
\end{equation}
(i.e., the maximum value of $\lambda_l$ for all $l$ such that
$\alpha_l > 0$) and
\begin{equation}
   \lambda_{\text{max}} = \text{min}(\lambda_l | \alpha_l < 0) .
\end{equation}
If $\lambda_{\text{max}} > \lambda_{\text{min}}$, then the line
segment (\ref{eq:line_ijk}) with $\lambda_{\text{min}} < \lambda <
\lambda_{\text{max}}$ forms an edge of a Voronoi polyhedron.  This
establishes the positions of two corners of the polyhedron:
\begin{eqnarray}
   \mathbf{x}_{c} & = & \mathbf{x}_{ijk} + \lambda_{\text{min}}
   \hat{\mathbf{n}}_{ijk} , \nonumber \\
   \mathbf{x}_{c'} & = & \mathbf{x}_{ijk} + \lambda_{\text{max}}
   \hat{\mathbf{n}}_{ijk} . \label{eq:xc}
\end{eqnarray}

The set of all nodes in the plane (\ref{eq:plane}) that lie closer to
the line segment $\lambda_{\text{min}} < \lambda <
\lambda_{\text{max}}$ than any other node defines what is called a
plaquette.  Since there are in general more than three such nodes, it
is convenient to define a unique label $q$ for each plaquette, with
\begin{equation}
   \hat{\mathbf{n}}_q \equiv \hat{\mathbf{n}}_{ijk} , \qquad
   \mathbf{x}_q \equiv \mathbf{x}_{ijk} ,
\end{equation}
for any members $\mathbf{x}_i$, $\mathbf{x}_j$, and $\mathbf{x}_k$ of
the given plaquette.  (The sign of $\hat{\mathbf{n}}_{q}$ is fixed by
some convention for the ordering of the nodes $\mathbf{x}_i$,
$\mathbf{x}_j$, and $\mathbf{x}_k$.)  Each plaquette is associated
uniquely with one edge of a Voronoi polyhedron, the length of which is
\begin{equation}
   d_{q} = |\mathbf{x}_{c'} - \mathbf{x}_{c}| = \lambda_{\text{max}} -
   \lambda_{\text{min}} , \label{eq:dq}
\end{equation}
with $d_q > 0$ by definition.

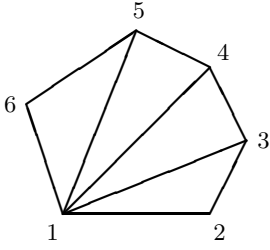
\begin{figure}
\setlength{\unitlength}{0.15em}
\begin{picture}(80,70)(0,0)
  \thicklines
  \put(20,10){\line(1,0){40}}
  \put(20,10){\line(5,2){50}}
  \put(20,10){\line(1,1){40}}
  \put(20,10){\line(2,5){20}}
  \put(20,10){\line(-1,3){10}}
  \put(60,10){\line(1,2){10}}
  \put(70,30){\line(-1,2){10}}
  \put(60,50){\line(-2,1){20}}
  \put(10,40){\line(3,2){30}}
  \put(15.5,3){1}
  \put(61,3){2}
  \put(73,28){3}
  \put(62,52){4}
  \put(39,63.5){5}
  \put(4,38){6}
\end{picture}
\caption{Partitioning of the surface $S_{ij}$ into triangles for the
area calculation in Eq.\ (\ref{eq:Sij}).}
\label{fig:Sij}
\end{figure}

At this point, one has sufficient information to determine whether a
link exists between any pair of nodes $\mathbf{x}_{i}$ and
$\mathbf{x}_{j}$.  The first step is to use the above procedure to
find all of the corner points $\mathbf{x}_c$ common to nodes $i$ and
$j$.  By definition, all such points lie in the plane
(\ref{eq:plane_ij}).  The set of these points defines a polygon, the
perimeter of which consists of the line segments (\ref{eq:dq}).  The
area of the polygon may be calculated by numbering the corner points
$\mathbf{x}_c$ in sequential order around the perimeter of the
polygon, then partitioning the polygon into triangles as shown in
Fig.~\ref{fig:Sij}.  
The normal area vector is
\begin{equation}
   \mathbf{S}_{ij} = \frac12 \sum_{c=3}^{N_{ij}} (\mathbf{x}_{c-1} -
   \mathbf{x}_1) \times (\mathbf{x}_{c} - \mathbf{x}_1) ,
   \label{eq:Sij}
\end{equation}
where $N_{ij}$ is the number of corner points common to nodes $i$ and
$j$.  The area $S_{ij} = |\mathbf{S}_{ij}|$ is the area of the surface
shared by the Voronoi polyhedra for sites $i$ and $j$; note that if
$N_{ij} = 1$ or 2, the shared region is a point or line, and the area
(\ref{eq:Sij}) is zero.  Nodes $\mathbf{x}_i$ and $\mathbf{x}_j$ are
linked only if $S_{ij} > 0$.

The volume $\Omega_i$ of the Voronoi polyhedron for node
$\mathbf{x}_i$ may be calculated from $S_{ij}$ and $d_{ij}$.  One
simply integrates the identity $\nabla \cdot \mathbf{x} = 3$ over the
polyhedron, using the divergence theorem and the fact that the plane
containing $S_{ij}$ is the perpendicular bisector of $\mathbf{d}_{ij}$
(although $\mathbf{d}_{ij}$ need not intersect $S_{ij}$ itself).  The
result is
\begin{equation}
   \Omega_{i} = \frac{1}{6} \sum_{j} S_{ij} d_{ij} .
\end{equation}
For each link $\ell \equiv (i,j)$, one can construct a polyhedron by
drawing lines from nodes $\mathbf{x}_i$ and $\mathbf{x}_j$ to each of
their common corner points $\mathbf{x}_c$.  The volume of this
polyhedron is
\begin{equation}
   \Omega_{\ell} = \frac{1}{3} S_{\ell} d_{\ell} .
\end{equation}
The volume $\Omega_{\ell} \equiv \Omega_{ij}$ is bisected by $S_{\ell}
\equiv S_{ij}$, with half lying in $\Omega_i$ and half in $\Omega_j$;
hence
\begin{equation}
   \Omega_i = \frac12 \sum_{j} \Omega_{ij} . \label{eq:split}
\end{equation}

The nodes in plaquette $q$ define a polygon in the plane
(\ref{eq:plane}); the perimeter of this polygon is formed by the links
$d_{\ell}$.  Hence, the area of the plaquette can be calculated in the
same way as the link area (\ref{eq:Sij}):
\begin{equation}
   \mathbf{S}_{q} = \frac12 \sum_{i=3}^{N_q} (\mathbf{x}_{i-1} -
   \mathbf{x}_1) \times (\mathbf{x}_{i} - \mathbf{x}_1) ,
   \label{eq:Sq}
\end{equation}
where $N_q$ is the number of nodes in plaquette $q$.  A polyhedron may
be constructed for each plaquette by drawing lines from each node of
the plaquette to the corner points (\ref{eq:xc}); the volume of this
polyhedron is
\begin{equation}
   \Omega_{q} = \frac{1}{3} S_{q} d_{q} . \label{eq:omegaq}
\end{equation}

Finally, the plaquette surfaces $S_q$ partition all of space into
nonoverlapping polyhedra (this is referred to as a Delaunay
tessellation \cite{Okabe00}).  These polyhedra (or {\em cells}) are in
one-to-one correspondence with the corner points $\mathbf{x}_c$ of the
Voronoi polyhedra.  The volume of cell $c$ is
\begin{equation}
   \Omega_c = \frac{1}{3} \sum_{q \in c} \mathbf{S}_q \cdot
   (\mathbf{x}_q - \mathbf{x}_c) , \label{eq:omegac}
\end{equation}
where the direction of $\mathbf{S}_q$ is chosen to point outward from
$\Omega_c$ (note that $\mathbf{x}_c$ does not necessarily lie inside
$\Omega_c$, \cite{ChrFriLee82a} so the dot product may be negative for
some $q$).

A useful set of sum rules for verifying the consistency of a
calculated geometry is
\begin{equation}
   \sum_i \Omega_i = \sum_{\ell} \Omega_{\ell} = \sum_{q} \Omega_q =
   \sum_{c} \Omega_c = \Omega ,
\end{equation}
where $\Omega$ is the volume of some region over which the node
distribution is periodic, such as a primitive cell in a Bravais
lattice (not to be confused with the generally nonperiodic cell
$\Omega_c$).  The sum rule for $\Omega_i$ follows directly from the
definition of $\Omega_i$ given in Sec.\ \ref{sec:topology}, since
every point in $\Omega$ must lie in at least one Voronoi polyhedron,
and the only regions of overlap between polyhedra are points, lines,
or planes of zero volume.  The sum rule for $\Omega_c$ was proven in
Ref.\ \onlinecite{ChrFriLee82a}.  The sum rule for $\Omega_{\ell}$
follows from that for $\Omega_i$, since the set $\{ \Omega_{\ell} \}$
is just another way of partitioning the set $\{ \Omega_i \}$ [see Eq.\
(\ref{eq:split})].  Likewise, the sum rule for $\Omega_{q}$ follows
from that for $\Omega_c$, since the set $\{ \Omega_{q} \}$ is just
another way of partitioning the set $\{ \Omega_c \}$ [see Eqs.\
(\ref{eq:dq}), (\ref{eq:omegaq}), and (\ref{eq:omegac})].

\section{Examples of link geometry}

\label{app:link}

This appendix presents values of the link lengths $d_{\ell}$ and
surface areas $S_{\ell}$ for several lattices.  The simplest geometry
occurs for Bravais lattices, of which only the cubic lattices are
considered here.  For the simple cubic lattice, only nearest neighbors
are linked, with $d_1 = a$ and $S_1 = a^2$, where $a$ is the lattice
constant.  For the body-centered cubic lattice, both first and second
nearest neighbors are linked, with
\begin{equation}
   d_1 = \frac{\sqrt{3}}{2} a , \qquad S_1 = \frac{3 \sqrt{3}}{16} a^2
   ,
\end{equation}
and
\begin{equation}
   d_2 = a , \qquad S_2 = \frac{1}{8} a^2 .
\end{equation}
For the face-centered cubic lattice, only nearest neighbors are
linked, with
\begin{equation}
   d_1 = \frac{a}{\sqrt{2}} , \qquad S_1 = \frac{a^2}{4\sqrt{2}} .
\end{equation}

The remaining lattices to be considered are those obtained by putting
symmetrized orbitals on the atomic sites of the diamond or zinc-blende
structure.  If only a single $s$ orbital per atom is used (i.e., one
$|000\rangle$ basis ket per atom), then each atom is linked to 4
nearest neighbors and 12 second-nearest neighbors, with
\begin{equation}
   d_1 = \sqrt{3} a , \qquad S_1 = 3 \sqrt{3} a^2,
\end{equation}
and
\begin{equation}
   d_2 = 2 \sqrt{2} a , \qquad S_2 = \frac{\sqrt{2}}{4} a^2 .
\end{equation}
Here $a = \frac14 a_0$, where $a_0$ is the conventional cubic lattice
constant.  Note that in this case, the link $d_2$ does not intersect
the surface $S_2$.

Coupling between second-nearest neighbors persists in models with more
than one orbital per atom.  In the ``$sp^3$'' model with four
$|111\rangle$ sites per atom (generated by applying the symmetry
operations of $T_d$ to $|r,r,r\rangle$ and $|a-r,a-r,a-r\rangle$),
each site is linked to three others on the same atom:
\begin{equation}
   d_0 = 2 \sqrt{2} r , \qquad S_0 = \frac{7 \sqrt{2}}{4} a^2,
\end{equation}
one on a neighboring atom:
\begin{equation}
   d_1 = \sqrt{3} (a - 2r) , \qquad S_1 = 3 \sqrt{3} a^2,
\end{equation}
and three second-nearest neighbors:
\begin{equation}
   d_2 = 2 \sqrt{2} (a - r) , \qquad S_2 = \frac{\sqrt{2}}{4} a^2 .
\end{equation}

In the model generated by either $T_d$ or $O_h$ and $|r,0,0\rangle$,
each of the six sites is linked to four others on the same atom:
\begin{equation}
   d_0 = \sqrt{2} r , \qquad S_0 = \frac{a^2(5a-6r)}{4\sqrt{2}(a-r)},
\end{equation}
four nearest neighbors:
\begin{equation}
   d_1 = \sqrt{3a^2 - 4ar + 2r^2} , \qquad S_1 = \frac{a^2
   d_1}{2(a-r)} ,
\end{equation}
and four second-nearest neighbors:
\begin{equation}
   d_2 = \sqrt{2} (2a - r) , \qquad S_2 =
   \frac{a^2(a-2r)}{4\sqrt{2}(a-r)}.
\end{equation}

In the model generated by either $T_d$ or $O_h$ and $|r,r,0\rangle$,
each of the 12 sites is linked to four others on the same atom, two of
which have
\begin{equation}
   d_0 = \sqrt{2} r , \qquad S_0' = \frac{a^2(3a-4r)}{4\sqrt{2}(a-r)},
\end{equation}
and two of which have
\begin{equation}
   d_0 = \sqrt{2} r , \qquad S_0'' = \frac{a^2(7a-8r)}{4\sqrt{2}(a-r)}.
\end{equation}
Each site is also linked to two nearest neighbors:
\begin{equation}
   d_1 = \sqrt{3a^2 - 8ar + 6r^2} , \qquad S_1 = \frac{a^2
   d_1}{2(a-r)} ,
\end{equation}
and one second-nearest neighbor:
\begin{equation}
   d_2 = 2\sqrt{2} (a - r) , \qquad S_2 = \frac{a^3}{2\sqrt{2}(a-r)}.
\end{equation}

In the model generated by $O_h$ and $|r,r,r\rangle$, there are two
distinct lattice sites.  Sites such as $|r,r,r\rangle$ and
$|a-r,a-r,a-r\rangle$ are labeled $b$ because they lie on the bonds
between atoms, whereas sites such as $|-r,-r,-r\rangle$ and
$|a+r,a+r,a+r\rangle$ are labeled $e$ because they lie on ``empty''
bonds.  (Both of these sites are actually Wyckoff $e$ sites, but they
are inequivalent, because the site symmetry of the atoms in diamond is
$T_d$.)  Each $b$ site is linked to three $e$ sites on the same atom
(and vice versa), with
\begin{equation}
   d_0^{be} = 2r , \qquad S_0^{be} = \frac{a (9a - 4r)}{8} .
\end{equation}
Each $b$ site is linked to one nearest-neighbor $b$ site:
\begin{equation}
   d_1^{bb} = \sqrt{3}(a-2r) , \qquad S_1^{bb} = \frac{3 \sqrt{3}}{4}
   a^2 ,
\end{equation}
and three nearest-neighbor $e$ sites:
\begin{equation}
   d_1^{be} = \sqrt{a^2 + 2(a-2r)^2} , \qquad S_1^{be} = \frac{1}{8} a
   d_1^{be} .  \label{eq:Sbe}
\end{equation}
Each $e$ site is also linked to three $b$ sites via (\ref{eq:Sbe}),
to six nearest-neighbor $e$ sites via
\begin{equation}
   d_1^{ee} = \sqrt{2a^2 + (a-2r)^2} , \qquad S_1^{ee} = \frac{1}{4} a
   d_1^{ee} ,
\end{equation}
and to three second-nearest neighbor $e$ sites via
\begin{equation}
   d_2^{ee} = 2\sqrt{2}(a-r) , \qquad S_2^{ee} = \frac{\sqrt{2}}{4}
   a(a+2r) .
\end{equation}


\begin{thebibliography}{81}
\bibitem[{not({\natexlab{b}})}]{note:monopoles}
\bibinfo{note}{Magnetic monopoles could be included in this formalism by paying
  close attention to phase changes of $2\pi n$ around a closed path.
  \cite{Dirac31,Rothe97,DeGrTous80} Such phase changes have been swept under
  the carpet here because magnetic monopoles are ordinarily of no interest in
  tight-binding theory. An alternative approach to magnetic monopoles is given
  in Ref.\ \onlinecite{WuYang75}.}
\bibitem[{not({\natexlab{c}})}]{note:charge}
\bibinfo{note}{Equation (\ref{eq:charge}) may also be derived from Eq.\
  (\ref{eq:Schr}).}
\bibitem[{not({\natexlab{d}})}]{note:trace}
\bibinfo{note}{As noted by Graf and Vogl, \cite{GrafVogl95} the relation
  $[x^{\alpha}, p^{\beta}] = i \hbar \delta_{\alpha\beta}$ can never be
  satisfied in a finite basis, because $\text{tr}(AB) = \text{tr}(BA)$ in a
  finite basis.}
\bibitem[{not({\natexlab{e}})}]{note:commute}
\bibinfo{note}{If the lattice happens to be a Bravais lattice, however, then
  one can show that \begin{displaymath} \sum_{j} \langle \mathbf{X}_i |
  [x^{\alpha}, p^{\beta} ] | \mathbf{X}_{j} \rangle \Omega_{j} = \frac{i
  \hbar}{2 \Omega_{i}} \sum_{j} d_{ij}^{\alpha} S_{ij}^{\beta} = i \hbar
  \delta_{\alpha \beta} , \end{displaymath} where the second equality follows
  from Eq.\ ($10 \frac12$) of Ref.\ \onlinecite{ChrFriLee82c} and the fact that
  a Bravais lattice has a center of inversion at the midpoint between two
  lattice sites, \cite{BurGla90} so that the center of mass of the surface
  $S_{ij}$ is just $\frac12 (\mathbf{x}_i + \mathbf{x}_j)$.}
\end{thebibliography}

\end{document}